\documentclass{aastex63}
\usepackage[utf8]{inputenc}
\usepackage{graphicx} % Required for inserting images

\begin{document}

\newcommand{\three}{$N\hspace{-0.2em}\ge\hspace{-0.2em}3$ }
\newcommand{\four}{$N\hspace{-0.2em}\ge\hspace{-0.2em}4$ }

\title{Architecture Classification for Extrasolar Planetary Systems}

\author[0000-0002-4884-7150]{Alex R. Howe}
\affiliation{The Catholic University of America, 620 Michigan Ave., N.E. Washington, DC 20064}
\affiliation{NASA Goddard Space Flight Center, 8800 Greenbelt Rd, Greenbelt, MD 20771, USA}
\affiliation{Center for Research and Exploration in Space Science and Technology, NASA/GSFC, Greenbelt, MD 20771}

\author[0000-0002-7733-4522]{Juliette C. Becker}
\affiliation{Department of Astronomy, University of Wisconsin-Madison, 475 N. Charter Street, Madison, WI 53706, US}

\author{Christopher C. Stark}
\affiliation{NASA Goddard Space Flight Center, 8800 Greenbelt Rd, Greenbelt, MD 20771, USA}

\author{Fred C. Adams}
\affiliation{Department of Physics, University of Michigan, 450 Church St, Ann Arbor, MI 48109}
\affiliation{Department of Astronomy, University of Michigan, Ann Arbor, MI 48109}

%\date{October 2024}

%\maketitle

% Note: we're maintaining the "peas-in-a-pod" terminology for now, but we reserve the right to change it in the final version.

\begin{abstract}

This paper presents a classification framework for the architectures of planetary systems based on a complete survey of the confirmed exoplanet population. With nearly 6000 confirmed exoplanets discovered, including more than 300 multiplanet systems with \three planets, the current observational sample has reached the point where it is both feasible and useful to build a classification system that divides the observed population into meaningful categories. This framework provides a criterion to split planetary systems into inner and outer regimes, and then further divides inner systems into dynamical classes. The resulting categories include ``peas-in-a-pod systems'' with uniformly small planets and ``warm Jupiter systems'' with a mix of large and small planets, as well as ``closely-spaced systems'' and ``gapped systems,'' with further subdivisions based on the locations of gaps and other features. These categories can classify nearly all of the confirmed \three systems with minimal ambiguity. We qualitatively examine the relative prevalence of each type of system, subject to observational selection effects, as well as other notable features such as the presence of hot Jupiters. A small number of outlier systems are also discussed. Potential additional classes of systems yet to be discovered are proposed.

% Omitted unless we have a good reason:
% Implications for planet formation and future surveys for exoplanets are discussed.

\end{abstract}


\section{Introduction}
\label{sec:intro}

At the present time, nearly 6000 planets outside our Solar System have been discovered with a reasonable degree of confidence. Significant efforts have been made to classify the planets themselves in recent years, usually based on size and orbital period (see, e.g., \citealt{Kopparapu18}). On the other hand, analogous work on the classification of planetary systems as distinct astronomical objects has been limited and has not yet considered the full population of discovered exoplanets in depth.

Planetary systems can display a wide range of different architectures, both in theoretical parameter space and in the currently observed sample. Over the past two decades, the number of observed multiplanet systems has grown to $\sim$1000, with more than 300 of them containing three or more planets. Although more data and more systems would be desirable, the database has reached a point where it becomes useful to provide a classification scheme for observed planetary systems. The goal of this paper is to identify meaningful classes of systems and provide a working version of such a classification scheme, based on a complete survey of the data from the NASA Exoplanet Archive \citep{Archive}.

Previous literature regarding architectures has explored one subset of planetary systems in depth: the so-called ``peas in a pod'' systems, in which several small (sub-Neptune-sized) planets of statistically similar sizes orbit close to their host star with regular logarithmic spacing \citep{Weiss18, Weiss23}. These peas-in-a-pod systems do indeed comprise a majority of known high-multiplicity systems. However, the definitions used for them vary \citep[e.g.][]{Zhuplus2018, Millholland21,Otegi22,Weiss23}, and they do not address systems that might be exceptions to the category. These exceptions are at most analyzed on a continuum with metrics such as dispersions of radii and period ratios. As part of our classification scheme, we endeavor to define reasonable bounds for the ``peas-in-a-pod'' class that can integrate consistently with the definitions of other classes.

Other explorations into planetary system architectures tend to focus either on types of individual planets that have particular dynamical influences on their systems \citep{Wang17}, or on other subpopulations such as systems hosting Earth-like planets \citep{Davoult}. Other studies have investigated more specific questions such as stability \citep{Volk} or the presence of outer planets \citep{HeWeiss}. However, none of the extant studies provide a classification scheme that takes into account the entire population of observed planetary systems.

Two frameworks have previously been proposed to characterize the full parameter space of planetary systems, specifically those proposed by \cite{GF20} and by \cite{Mishra22,MishraII}. \cite{GF20} took an information-theoretic approach and computed various statistical measures for a sample of planetary systems drawn from the California Kepler Survey \citep{CKS} with small enough error bars, comprising 206 two-planet systems and 129 \three systems. The final set of statistics they selected included star-planet mass ratio, mass partitioning, dynamical spacing, gap complexity, and variance in inclination.

\cite{GF20} ran a cluster analysis on these statistics to identify natural categories of planetary system architectures. However, this analysis resulted in only two clusters, defined by a gap complexity of $\mathcal{C}>0.33$ and $\mathcal{C}<0.33$. They hypothesized that the high gap complexity cluster was not physical, but was caused by undetected planets creating apparent gaps in the dynamical spacing. (Such a large gap complexity effectively requires one gap to be several times wider than the others in logarithmic space, with a period ratio $>$10 in most cases. However, some instances of a divide between inner and outer planets as we define in Section \ref{sec:io} may support a physical gap complexity this large.)

The second proposed framework, by \cite{Mishra22}, took a more theoretical approach. They defined four classes of system architectures \textit{a priori} based on first-order approximations to the possible parameter space of mass ordering. (They did not include period spacing in this framework, although they note that the same \textit{a priori} classes can be applied to other parameters.) Their four classes were ``ordered'' (that is, increasing in mass with distance from the star), ``anti-ordered,'' ``mixed,'' and ``similar'' (this last corresponding to the peas-in-a-pod case).

\cite{Mishra22} verified this framework by comparing it with both population synthesis models and observed planetary systems, in this case choosing a sample of 41 systems with \four measured masses. Population synthesis models showed significant clustering around all four classes, and they also showed emergent formation pathways that could generate each class \citep{MishraII}. However, the observed systems were overwhelmingly classified as ``ordered'' or ``similar,'' with only two (GJ 876 and Kepler-89) classified as ``mixed'' and none at all classified as ``anti-ordered.'' This suggests that the mass ordering of planets may be significant, but some refinement of the categories is needed. Additionally, this framework did not address period spacing, as \cite{GF20} did, and which has been shown to be predictive in later analyses (e.g. \citealt{HeWeiss}).

In this paper, we present a novel classification scheme for planetary system architectures based on a complete survey of the NASA Exoplanet Archive catalog (more specifically, the Planetary Systems Composite Parameters Table \citep{Archive}, which lists all confirmed exoplanets, ideally with their best available measured parameters). In addition to being the first proposed classification to directly consider the entire catalog, and thus to include planets from all detectable parts of the parameter space, we also consider both mass (and mass ordering) and period spacing when designing our framework. This framework allows us to make a partial classification of one- and two-planet systems and a nearly complete classification of known systems with three or more planets, with a very few exceptions with unusual dynamical structures (see Section \ref{sec:outliers}). The wide applicability of this framework indicates that exoplanetary systems do not appear as diverse as is usually thought, and our own Solar System is not as unusual as is usually thought, but can fit reasonably well in the standard ``peas-in-a-pod'' category under our definition.

\begin{figure}[!ht]
    \centering
    %\hspace{0.55in}
    \includegraphics[width=0.65\textwidth]{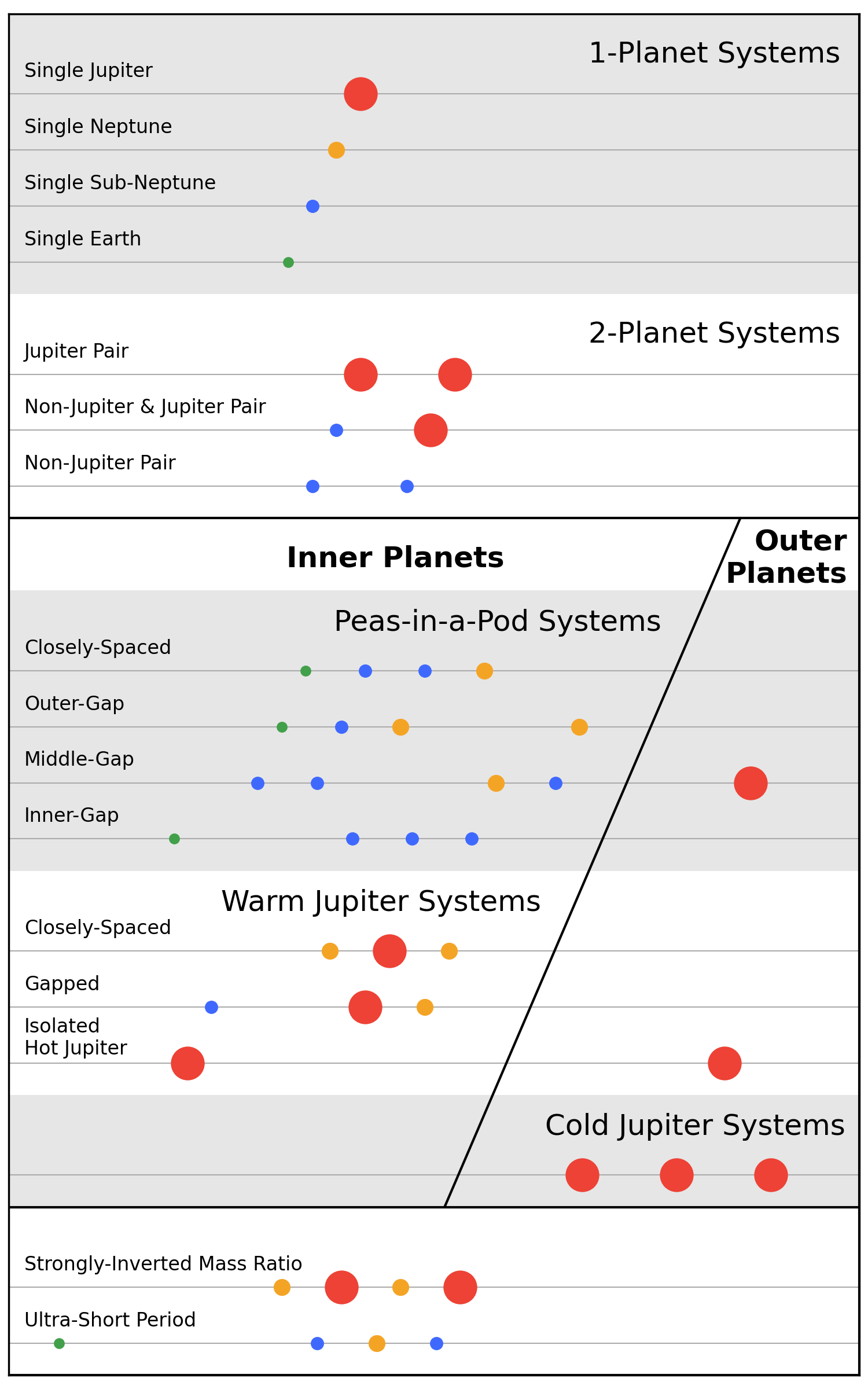}
    \caption{Quick-reference chart for our classification of planetary system architectures, with representative model systems for each category. Each row corresponds to one planetary system, with horizontal spacing corresponding to orbital period on a log scale and point sizes corresponding to planet size. Colors correspond to planet type, as described in Section \ref{sec:classes}: Jupiters ($>$6 $R_\oplus$, red), Neptunes (3.5-6 $R_\oplus$, gold), Sub-Neptunes (1.75-3.5 $R_\oplus$, blue), and Earths ($<$1.75 $R_\oplus$, green). (However, the exact composition of non-Jupiters in this figure is arbitrary.) This format is used for all plots of planetary system architectures throughout this paper. \\
    \textbf{Framework summary:} Each system is divided into inner and outer planets (if both are detected). Systems with \three inner planets are classified based on whether their inner planets include any Jupiters, and whether (and if so where) their inner planets include large gaps with a period ratio $>$5. Other dynamical features are addressed separately from the overall classification system (see Section \ref{sec:other}).}
    \label{fig:classes}
\end{figure}

The core of our classification system comes down to three questions for any given system (although in several cases we add additional subcategories). Does the system have distinct inner and outer planets? Do the inner planets include one or more Jupiters? And do the inner planets contain any gaps with a period ratio greater than 5? We find that these three questions are sufficient to classify $\sim$97\% of multiplanet systems with \three planets with minimal ambiguity, to which we then add useful subcategories, such as where any large gaps occur and whether or not a hot Jupiter is present. (Hot Jupiters can also be considered as their own class, although a small amount of overlap exists.) Figure \ref{fig:classes} illustrates the basic elements of this framework, with greater detail provided in the corresponding sections for each category.

Given that this framework is based on a catalog of all confirmed planets, with only minimal filtering for those that have suitable host stars and usable measurements, it will be subject to many observational biases, including combining different survey methods, initial target selection, follow-up target selection, and actual detectability of planets with current methods. For example, transit surveys are sensitive to different parts of the parameter space than radial velocity surveys, which are different still from other detection methods. Additionally, while the number of observed systems is large enough to define a classification scheme, there are not enough systems in most of the categories to do meaningful statistics. As a result, any statistical claims we make are necessarily qualitative. Some trends do appear clearly at the $\sim$10\% level (see especially Table \ref{tab:all_inner}), but we do not attempt any rigorous statistical analysis, as such is beyond the scope of this paper and would likely require a larger population sample.

It is also possible that some systems could move from one class to another due to the discovery of additional planets, for example, moving from a ``gapped'' system to a closely-spaced system due to the discovery of a planet in the gap. Any ``missing'' planets will also not be drawn from a uniform sample, but would be preferentially drawn from the parts of the parameter space where current surveys are not sensitive. However, as we discuss in Section \ref{sec:inner}, the available observational evidence suggests that the \textit{existence} of the categories we identify is robust in aggregate, even if we do not have a clear idea of their extents, and the placement of individual systems can be uncertain.

The primary goal of this paper is to establish the classification scheme itself, one that is as broad in scope as possible and into which future discoveries, even unexpected ones, can be placed reliably. That these categories can be defined at all is, we believe, a robust conclusion despite the large observational biases in the sample.

This framework also provides an alternative to population-synthesis-derived population models for use in simulations of exoplanet populations for future observations \citep[e.g.][]{Howe24}. This could allow for more physically motivated and observationally supported simulated populations than occurrence rates or population synthesis models alone.

This paper is organized as follows. In Section \ref{sec:filter}, we discuss our data selection, filtering, methods, and our definitions of individual planet types. Sections \ref{sec:1pl} and \ref{sec:2pl} address the classification of one-planet systems and two-planet systems, respectively. Section \ref{sec:3plus} addresses higher-multiplicity systems and presents the main part of our classification framework. Section \ref{sec:other} examines other dynamical features that are not covered by our framework such as ``inverted'' mass ratios and hot Jupiters, as well as outlier systems. We discuss the limitations of our framework and potential categories of systems that have not yet been detected in Section \ref{sec:discuss}. Finally, we summarize our classification scheme and discuss its implications and potential future work in Section \ref{sec:conclusion}.

\section{Filtering the Database}
\label{sec:filter}

% Data for the 09/24/2024 pull from the catalog
% Planets listed 5759
% Planets accepted 5686
% Systems accepted 4259
% Transit discoveries: 4282
% RV discoveries: 1080
% Other: 324
% Microlensing: 211
% Imaging: 61
% Transit timing variations: 28
% Eclipse timing variations: 15
% Orbital brightness variations: 4
% Astrometry: 2
% Pulsar timing variations: 2
% Disk kinematics: 1

The NASA Exoplanet Archive lists all confirmed exoplanets in their Planetary Systems Composite Parameters Table \citep{Archive}. In the version of the database we used, accessed on September 24, 2024, 5759 planets were listed, which we later filtered down to 5686 as described in Section \ref{sec:exclude}. Of these planets, 75\% were initially discovered by transits, 19\% by radial velocities, and 6\% by other methods (most often microlensing). They range in size from Mercury-sized (0.31 R$_\oplus$ and 0.015 M$_\oplus$) to super-Jupiters at the brown dwarf limit (13 M$_J$) and young giant planets that have not yet cooled and retain radii as large as 3 R$_J$. Their orbital periods span more than nine orders of magnitude from 4.3 hours to 1.26 million years. Regarding host stars, we removed stellar remnants, brown dwarfs, and pre-Main Sequence stars, although we allowed giants to remain. The resulting sample of host stars range in mass from 0.08 to 11 M$_\odot$. Their spectral types range from B2 to M8, with most of them being G, K, or early M-type Main Sequence stars.

\subsection{Exclusions and Corrections}
\label{sec:exclude}

While the NASA Exoplanet Archive is a comprehensive catalog of confirmed planets, the data it includes are copied from discovery and characterization papers, which can at times be inaccurate, unreliable, poorly-constrained, contradicted by other papers, or copied incorrectly in spite of best efforts. Occasionally, this is true even of the \textit{existence} of listed planets. Additionally, a few planets such as pulsar planets are unsuitable for inclusion in our framework. Therefore, we filtered the catalog to remove non-usable data, or where possible to correct or recalculate it (most notably, recalculating upper bound mass and radius measurements based on the mass-radius relation of \citealt{Chen+Kipping}). This filtering removed fewer than 2\% of planets from the catalog, although a larger number ($\sim$5\%) had corrections or recalculations made.

We removed a number of planets and systems from the dataset that had insufficient data, unreliable data, or did not represent the ``normal'' planet population in some way. In the first step, we removed all planets listed as orbiting pulsars or white dwarfs because it is likely that these are either ``second generation'' planets (those that formed in a second phase of planet formation after the host star left the main sequence), or they have been significantly perturbed as the star evolved off the main sequence \citep[e.g.][]{Debes2002, Veras2021}. Either way, these planets do not represent the population of planets orbiting main sequence stars. We removed all planets orbiting brown dwarfs (based on a host mass of $<$0.08 M$_\odot$). We also removed all planets with insufficient data for categorization, specifically, those for which neither mass nor radius measurements were available, those for which neither period nor semi-major axis measurements were available, and those for which a stellar mass was not available (which makes it impossible to verify a valid host mass or to compute any missing period values).

% KMT-2020-BLG-0414L was removed because it is a white dwarf host, even though it is not listed as such in the catalog. There's no easy way to know if there are others, especially in the microlensing samples.

Several more stars were removed from the dataset due to other incomplete or unreliable data. HD 100546 was removed because its sole planet does not have a reliable mass or radius measurement due to the presence of a circumplanetary disk \citep{Quanz15}. Kepler-132 was removed because its planets are split between the two stars of an unresolved binary and thus cannot be assigned to specific planetary systems \citep{Rowe14}. Kepler-70 was removed because the existence of its planets is considered uncertain \citep{Blokesz19}, and in any case because they represent post-main sequence evolution. And V1298 Tau \citep{David19} and ITG 15 A \citep{Michel24} were removed because they represent pre-main sequence evolution, and their planets' masses cannot be reliably estimated.

These exclusions also removed a number of individual planets within systems. HD 105618c and HD 153557d were removed because they are distant brown dwarf companions without analogs among the confirmed planetary-mass population \citep{Feng22}. Additionally, while cross-checking the entire database was prohibitive, we paid special attention to high-multiplicity systems with five planets or more, as they are particularly high-interest for future observations and analysis. Most of these we included as reported. However, as a result of these checks, we removed a few additional planets that are disputed or considered to be low-probability solutions. This reduced HD 34445 from a six-planet solution to a one-planet solution, including only planet b \citep{Rosenthal21}; and it reduced GJ 667C from a five-planet solution to a two-planet solution, including only planets b and c \citep{Feroz14}.

% HD 219134: Planet g is just not mentioned in most of the followup papers, yes or no. Recovered with an aliased period in 2021AJ....161..134H
% HD 191939: Planet f is an outer Jupiter companion based on a long-term trend, but is robust to follow-up (2024arXiv240906795L)
% HD 40307: we used the Archive's 5-planet solution including the disputed planet g, but excluding the disputed planet e. 2016A&A...585A.134D detected the signal for planet g, but considered it doubtful.
% Kepler-122: I was concerned about bad TTV mass measurements, but that didn't end up mattering to our discussion.

We similarly checked the data for all of the systems we identified as potential outliers for disputed planets and unreliable data. This resulted in the removal of a number of other planets, including WASP-18c \citep{Cortes20} and WASP-126c \citep{Maciejewski20}, which otherwise would have been the only candidates for multiple hot Jupiters in a single system, and which illustrates the importance of careful analysis for such outlier cases.

% HD 10180: as many as 9 planet candidates proposed (2012A&A...543A..52T), but a 6-planet solution is currently accepted (2021A&A...645A...7K)
% HIP 41378: 5-planet solution widely accepted, including by our previous paper.
% HR 8799: 4 planets clearly visible by direct imaging.
% Kepler-42: followed-up and well-checked for false positives (2011ApJS..197....5F)
% Kepelr-51: densities obtained with TTVs, but at fairly high confidence (2014ApJ...783...53M)
% Kepler-90: 8-planet solution identified with machine learning, but seemingly well-accepted (2018AJ....155...94S)
% WASP-148: both planets confirmed by RV follow-up (2020A&A...640A..32H)

% Provisionally not citing discovery/characterization papers in this paragraph because it relates to how things are constrained in the archive itself, not papers disputing the archive.
Missing radius and mass measurements that did not result in a system being excluded entirely were computed from the \cite{Chen+Kipping} mass-radius relation. We did the same for any masses and radii that were listed as being limits rather than well-constrained measurements, with two exceptions. HD 219134d and f have only limits measured for both masses and radii. However, their well-confirmed status in an otherwise-better-constrained six-planet system led us to leave them as reported. We also used the \cite{Chen+Kipping} relation for nominally-constrained measurements that were not reliable enough to use: specifically, Kepler-33c and HD 93963b.

Some mass limits we set to 1 $M_J$. These were any radii $>$11.854 $R_\oplus$, for which the \cite{Chen+Kipping} relation is not single-valued, with the limiting radius based on the Exoplanet Archive's mass cutoff of 30 $M_J$. Likewise, any masses estimated (based on radii) by the Exoplanet Archive to be $>$30 $M_J$ were reset to 1 $M_J$ automatically. Since we applied measured radii with higher preference than measured or estimated masses in classifying planets, and since all of the planets affected were classified as Jupiters regardless (see Section \ref{sec:classes}), this did not significantly change our results.

Despite these efforts, it is possible that some errors remain in our adopted dataset. For example, KMT-2020-BLG-0414L is a known white dwarf host \citep{Zhang24}, but was not listed as such in the catalog. It is not practical to check the entire database for similar cases, nor is it practical to check for data that may be incorrect, but nevertheless pass a sanity check. However, the small number of exclusions we did make and the special attention we paid to high-interest systems suggest that any remaining errors will not have a significant impact on our results.

In total, we excluded 73 out of 5759 planets in the September 24, 2024 version of the NASA Exoplanet Archive. The end result of this filtering process was a reasonably well-supported list of 4259 planetary systems. We have listed the census of systems in our sample by multiplicity in Table \ref{tab:npl}.

\begin{table}[htb]
    \centering
    \begin{tabular}{l | r}
    \hline
    $N_{\rm{pl}}$ & Number of Systems \\
    \hline
    1 & 3311 \\
    2 & 634 \\
    3 & 202 \\
    4 & 74 \\
    5 & 26 \\
    6 & 10 \\
    7 & 1 \\
    8 & 1 \\
    3+ & 314 \\
    Total & 4259 \\
    \hline
    \end{tabular}
    \caption{Number of confirmed planetary systems by multiplicity after applying our filters (NASA Exoplanet Archive, \citealt{Archive}, accessed September 24, 2024).}
    \label{tab:npl}
\end{table}

\subsection{TTVs and Mass Filtering}

A more complex problem was that several dozen entries in the Exoplanet Archive have mass and radius measurements that result in unphysically high (or in one case low) densities, sometimes far higher than even a pure iron planet would be. These results are usually derived from transit timing variations (TTVs), which can yield implausibly large mass measurements, often with very large error bars.

A few of these unphysical results could be corrected by filling in data from better sources not used in the Exoplanet Archive's composite table. These included radius measurements for HATS-12b \citep{HATS-12}, K2-229b \citep{K2-229}, Kepler-297d, and Kepler-1979b (both \citealt{Kepler-297}), mass measurements for Kepler-444d and e \citep{Kepler-444}, and all of the data for the HIP 41378 planets \citep{HIP41378}. Additionally, we filled in stellar mass measurements for CT Cha \citep{CTCha} and HIP 77900 \citep{HIP77900}.

However, most of the TTV results could not be corroborated by other (usually RV) data. We dealt with this by recalculating any masses that led to unphysical densities higher than pure iron based on the \cite{Chen+Kipping} mass-radius relation. However, even after these corrections, it was not clear which TTV mass measurements were trustworthy, and some of the resulting mass ratios between planets are therefore equally suspect. (The large majority of the measurements in question derive from \cite{HaddenTTVs}, which did not apply strong priors to the compositions of planets measured by TTVs.) Therefore, we used a more complex criterion.

For radii $<$1.75 $R_\oplus$, we expect the planet to be likely bare rock (although the \cite{Chen+Kipping} relation switches to a gas-rich composition as early as 1.22 $R_\oplus$), so we need only worry about ensuring that the planet has a density less than that of pure iron. To do this, we set the maximum mass to be the same as the mass-radius relation for rocky planets scaled up by a factor of 8.0/5.5.

For radii $>$1.75 $R_\oplus$, we expect the planet to likely have a thick hydrogen atmosphere. Some of these planets may still be bare rock, especially ultra-short period planets (USPs), but the difference between an Earth-like composition or even a Mercury-like composition and pure iron provides some buffer to allow us to plausibly model a bare rock with a toy model of a pure iron core plus a relatively shallow hydrogen atmosphere. For the range of 1.75-2.25 $R_\oplus$, we set the maximum mass to be that of a 1.75 $R_\oplus$ sphere of pure iron.

For radii of 2.25-6.0 $R_\oplus$, we set the limiting mass to be that of a pure iron core 0.5 $R_\oplus$ smaller than the planet's observed radius. This model does not account for the potentially larger mass of the atmosphere, again providing some buffer for our toy model. In most cases, this will predict a core mass much larger than the true value. (For planets larger than 6 $R_\oplus$, we do not exclude any mass measurements.)

We also recalculated all of the masses for Kepler-92 and Kepler-238 because our criterion would have recalculated only some of the masses, and this produces implausible mass ratios. (All other such cases had expected terrestrial or gaseous planet densities.) A few entries were also deemed unphysical due to erroneous radii. This was notably the case for Kepler-37e, which does not have a published radius. For this planet, we recalculated the radius instead of the mass.

\subsection{Planet Classes}
\label{sec:classes}

Following from \cite{Kopparapu18}, we divided exoplanets into four classes based on their radii. These four classes are Jupiters ($>$6.0 $R_\oplus$), Neptunes (3.5-6.0 $R_\oplus$), sub-Neptunes (1.75-3.5 R$_\oplus$), and Earths ($<$1.75 $R_\oplus$). To better fit the observed population, we have merged the two smallest categories of super-Earths (1.0-1.75 R$_\oplus$) and Sub-Earths ($<1.0$ $R_\oplus$) from the \cite{Kopparapu18} grid given the small number of observed sub-Earths, and we have also removed the minimum and maximum radius limits (0.5 $R_\oplus$ and 14.3 $R_\oplus$, respectively). For planets for which measured radii are not available, we estimated their radii from measured masses based on their mass-radius relation of \cite{Chen+Kipping}. This relation sets break points between classes of 3.7, 12.0, and 30.1 $M_\oplus$.

These four classes of planet sizes are useful because they reflect natural break points in the observed planet population, especially in the \textit{Kepler} sample \citep{Fulton17}. which comprises more than half of the dataset. They are also physically motivated, with 1.75 $R_\oplus$ falling in the middle of the radius valley and 30 $M\oplus$ being near the mass scale for runaway accretion \citep{Pollack1996} and the transition to gas giants. In this paradigm, we find that the presence or absence of Jupiters under this definition is one of the larger factors in determining the architecture of a planetary system. Given the smaller population of Neptunes, it may be preferable to instead group the Sub-Neptunes, Neptunes, and even the smaller ``Jupiters'' into a single class ranging from 1.75 to 8 $R_\oplus$ (4 to 50 $M_\oplus$). However, for consistency, we retain the standard classes recognized by the literature. (Planets in the 6-8 $R_\oplus$ range are few enough that it does not make a large difference to the population as a whole.)

We manually adjusted the assigned classes for the so-called ``super-puffs,'' which are Neptune- or sub-Neptune-mass planets with unusually large radii and low densities. While a number of different definitions of super-puffs have been proposed \citep[e.g.][]{JH19,Lee2019,Wang2019,Gao2020,Piro2020}, we find that the best fit for our purposes is the one used by \cite{SuperPuffDef}, giving a density limit of $\rho<0.3$ g cm$^{-3}$, to which we add the constraint that they be lower-mass than our Jupiter cutoff of 30.1 $M_\oplus$. We classify all super-puffs as ``Neptunes'' regardless of our other criteria.

We also investigated the possibility of dividing the exoplanet population by period (or instellation) into ``hot,'' ``warm,'' and ``cold'' populations. This proved to be unnecessary, as the population is better described by empirical break points in the period (and period ratio) distribution. However, we did retain some common period-based definitions for subsets of the population, such as $<$10 days for hot Jupiters \citep{Wang15} and $<$1 day for ultra-short period planets (USPs, \citealt{Winn18}).

\section{Single-Planet Systems}
\label{sec:1pl}

\begin{figure}[!ht]
    %\centering
    %\hspace{0.55in}
    \includegraphics[width=0.99\textwidth]{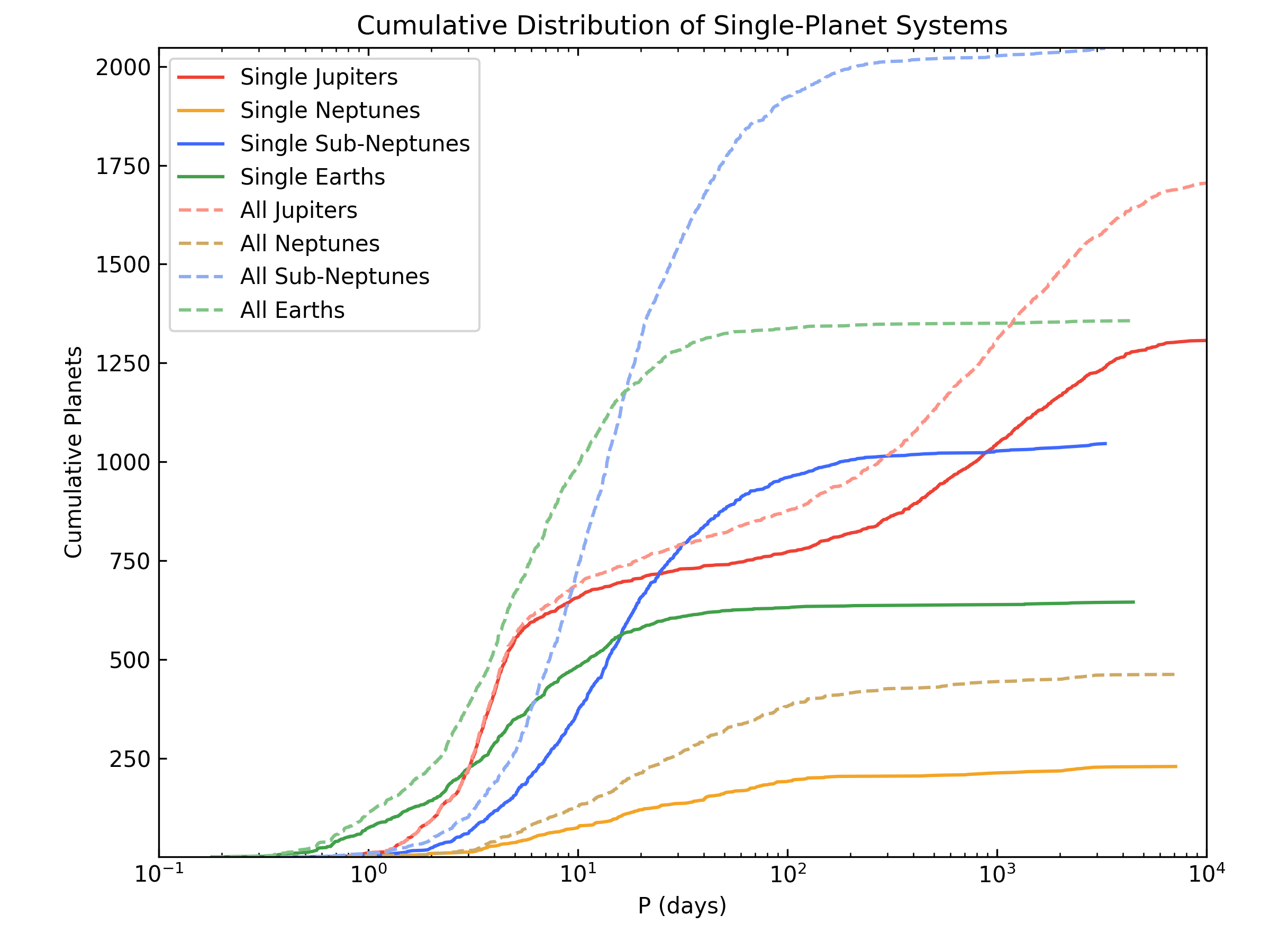}
    \caption{Cumulative distributions of confirmed exoplanets with period, comparing total numbers of planets (dashed) to those in single-planet systems (solid), according to the same color scheme as Figure \ref{fig:classes}. Hot Jupiters show far fewer companions than other planet types, as illustrated by the near-coincidence of the two Jupiter distributions at $<$10 days.}
    \label{fig:singles}
\end{figure}

Over three quarters of the planetary systems in our dataset (78\%) have only a single confirmed planet. While these cannot be classified in the same sense in terms of architectures, we can draw some conclusions by comparing them with multiplanet systems. Many or perhaps even most of these systems will have a single known planet because of selection effects. However, any planet types that preferentially occur \textit{without} companions, such as hot Jupiters, should be over-represented within this group. To analyze this possibility, we compare the cumulative distribution of each size class of planets over orbital period between single planets and all planets. The result of this comparison is shown in Figure \ref{fig:singles}.

As expected, Jupiter-sized planets are far less likely to occur in multiplanet systems at periods of $<$10 days and virtually none do at $<$5 days, as indicated by the near-coincidence of the two Jupiter distributions at those periods. Meanwhile, roughly half of all other planet types and even a third of Jupiters at periods $>$10 days occur in multiplanet systems.

It is plausible that many if not most of these single-planet systems harbor additional planets, which simply have not been found due to selection effects and incompleteness of surveys. However, much like the hot Jupiters, some $\sim$95\% of which do not have known companions and whose formation is expected to disrupt other planets, it is plausible that some unknown number are truly isolated planets -- the only ones in their systems. This could be the case for famous planets like GJ 436b and GJ 1214b, for which no companions have been found despite significant effort \citep[e.g.][respectively]{Lanotte14,Harpsoe13}. However, these systems could equally harbor smaller, Mars-sized planets or distant giant planets that are not detectable at all with current methods.

\section{Two-Planet Systems}
\label{sec:2pl}

We treat two-planet systems separately from those of higher multiplicity both because they show significantly different statistics, as we discuss in Section \ref{sec:3plusclass}, and because we can analyze them with different (and simpler) metrics. In particular, two-planet systems allow us to more easily compare period ratios and mass ratios of pairs of planets, since there is only one pair in each system. On the other hand, this means that our classification scheme for two-planet systems is necessarily less complete than for higher-multiplicity systems.

\begin{figure}
    \includegraphics[width=0.99\textwidth]{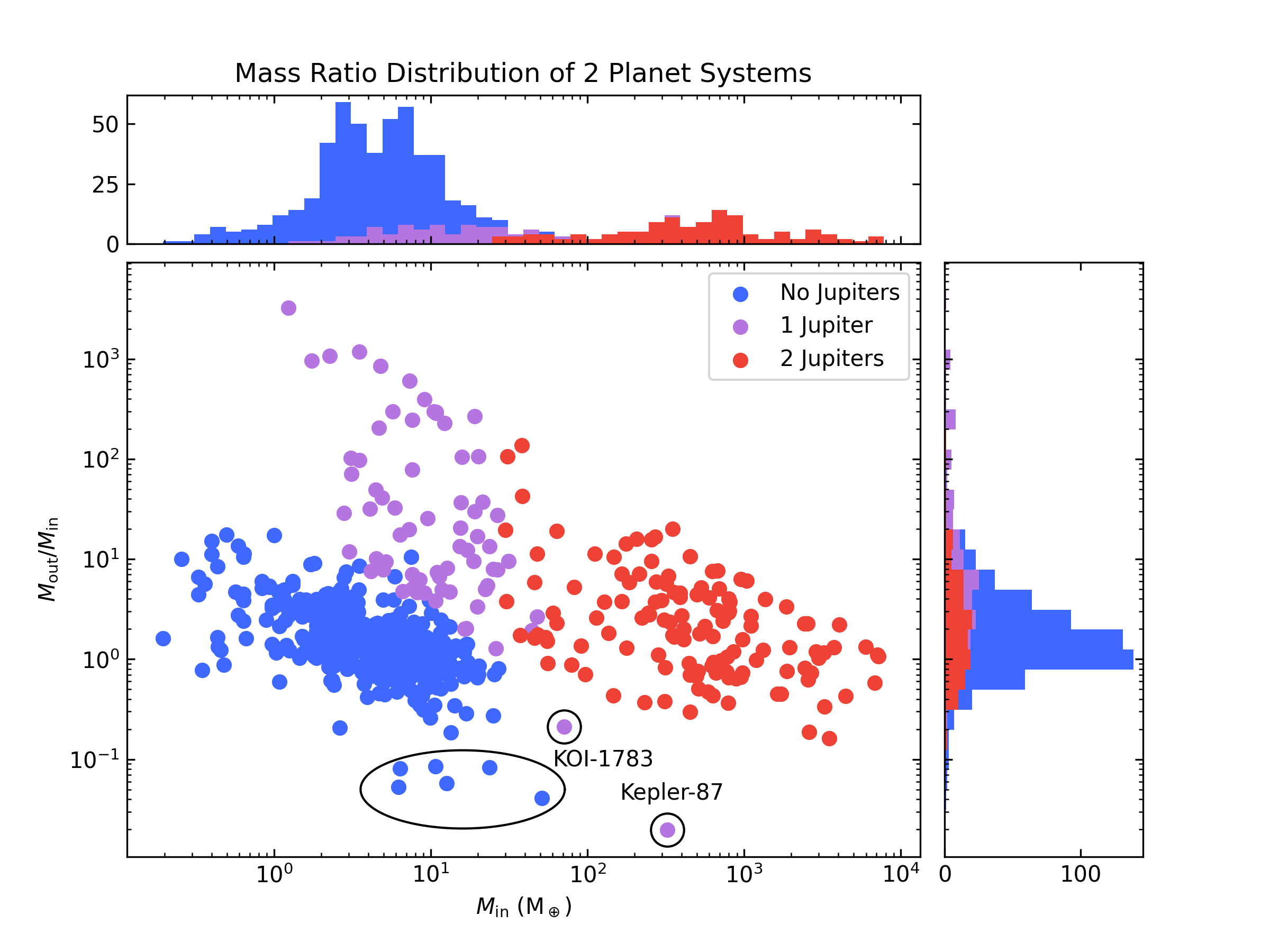}
    \caption{Scatter plot and histograms of two-planet systems by mass ratio versus inner planet mass, divided into Jupiter pairs (red) Jupiter and non-Jupiter pairs (lavender), and non-Jupiter pairs (blue, here representing all types of non-Jupiters). Overlaps between the color-coded regions (defined by planet size) are due to differences in assigning Jupiter status by radius rather than mass when possible. Systems with unusually low mass ratios (smaller outer planets) are highlighted.}
\label{fig:2planet_m}
\end{figure}

\begin{figure}
    \includegraphics[width=0.99\textwidth]{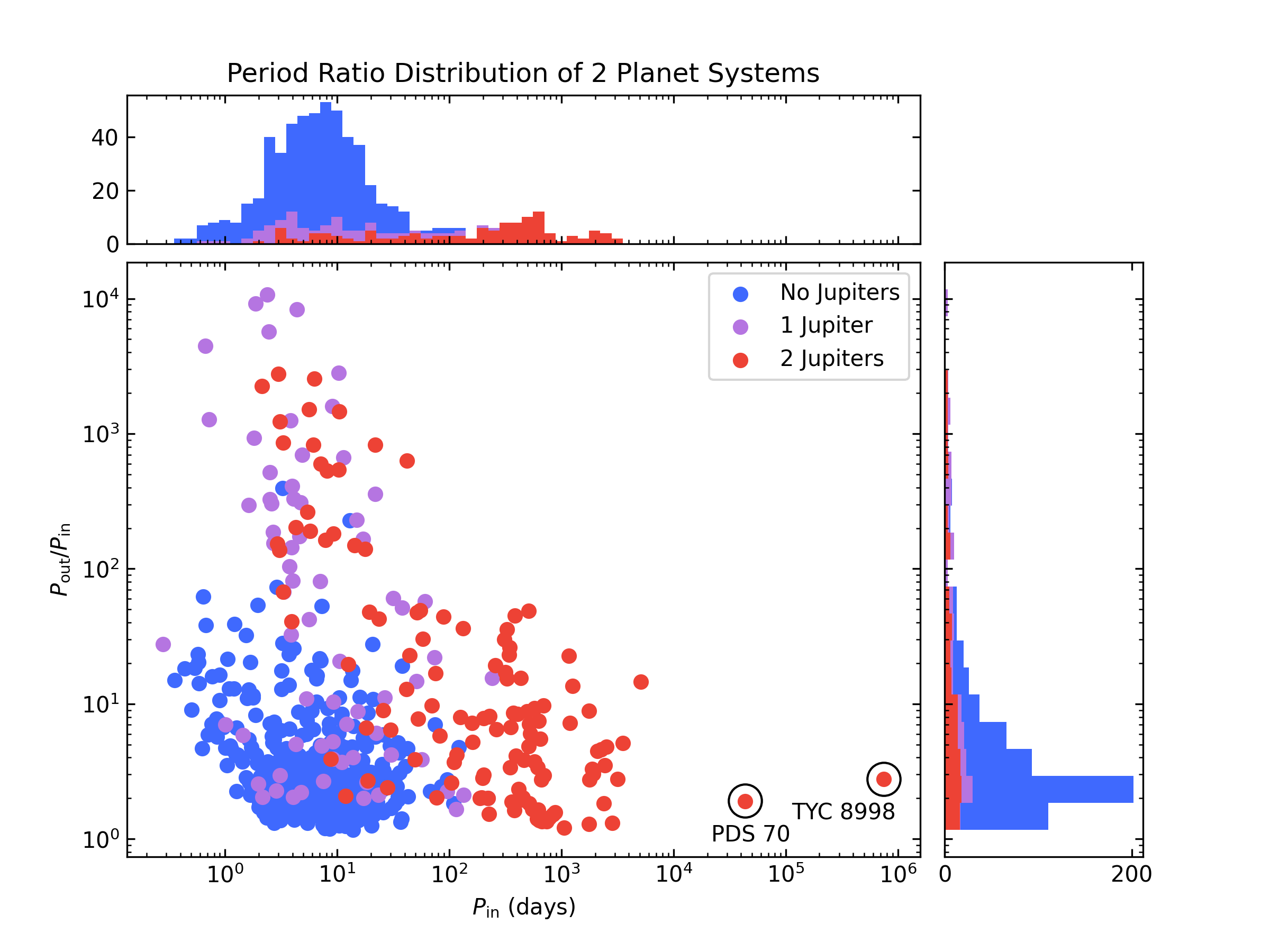}
    \caption{Same as Figure \ref{fig:2planet_m} for period ratio versus inner planet period. (Note that the color-coded regions defined by planet size overlap strongly in period space.) Jupiters show more systems with longer periods and larger period ratios. Outliers to the right of the plot correspond to directly-imaged systems.}
\label{fig:2planet_p}
\end{figure}

Scatter plots and histograms for two-planet systems in terms of mass ratios and period ratios are shown in Figures \ref{fig:2planet_m} and \ref{fig:2planet_p}, respectively. In these figures, we distinguish three classes of two-planet systems: those with no Jupiters in blue, those with exactly one Jupiter in lavender, and those that feature Jupiter pairs in red. These three classes, being defined by planet size, form nearly non-overlapping regions in Figure \ref{fig:2planet_m}. (They are strictly overlapping for measured masses, but not for masses inferred from radii.) However, there are also distinctive patterns in period ratio space, with long-period planets ($\gtrsim300$ days) being exclusively Jupiter pairs and large period ratios ($\gtrsim100$) almost exclusively including at least one Jupiter. This points to systematically longer periods for Jupiters than for small planets. This may be the result of both formation processes (with Jupiters being more common beyond the ice line) and detection biases (with Jupiters being much more detectable by radial velocities at long periods).

We also note that for systems with two non-Jupiters, both the period ratios and the mass ratios of the planet pairs show a well-defined and relatively narrow peak. The period ratios cluster around a value of $\sim$2, whereas the mass ratios cluster near unity, consistent with the ``peas-in-a-pod'' pattern found for higher-multiplicity systems \citep{Weiss18}. In contrast, the two planet systems containing one (or two) Jovian planets display a much wider range of period ratios and especially mass ratios, with the histograms relatively flat over the ranges displayed.

Three systems in Figure \ref{fig:2planet_p} appear to be outliers for their classes (although not for the population as a whole). However, each of these can be largely explained by other factors. We classified K2-141 (the lavender dot at the far left) as a Jupiter-Earth pair, but the radius of K2-141c is poorly measured due to grazing transits and the lack of a detected RV signal, and it may be a Neptune-Earth pair by mass \citep{K2-141}. We also classified HD 134060 (the highest blue dot) as a no-Jupiter system, but it can be better analyzed as a one-Jupiter system, as HD 134060c is just below our Jupiter-Neptune cutoff at 29.3 $M_\oplus$ \citep{HD134060}.

The case of GJ 411 (Lalande 21185, the second-highest blue dot) is more complex. It is not an outlier in terms of confirmed planets. However, its status as the fourth-closest star system to the Sun at only 2.5 pc and its status as an M-dwarf (thus giving it high detectability for transits and radial velocities) makes it a potential high priority for more detailed study, so it may be an outlier due to observational biases. Additionally, the system is suspected to have an unconfirmed third planet, which would move it out of the two-planet category entirely \citep{GJ411}.

\begin{figure}[!ht]
    %\centering
    %\hspace{0.55in}
    \includegraphics[width=0.99\textwidth]{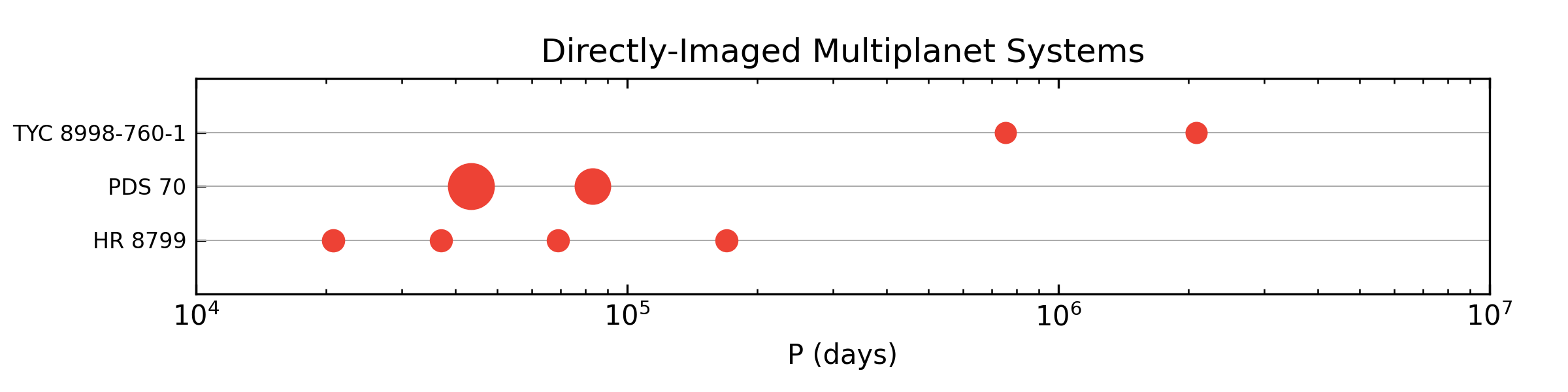}
    \caption{Architectures of directly-imaged multiplanet systems, using the same format as Figure \ref{fig:classes}. While not a separate category of systems themselves, direct imaging is sensitive to planets at very wide separations, so that these systems are dramatic outliers from the rest of the currently observed population. All known directly-imaged planets are Jupiters.}
    \label{fig:direct}
\end{figure}

Figure \ref{fig:2planet_p} does show two clear outliers in PDS 70 and TYC 8998-760-1. These are two of only three known directly-imaged multiplanet systems, the other being HR 8799 (see Figure \ref{fig:direct}). The rarity of these systems is owed to the relatively small number of directly-imaged planets overall and the unlikelihood of multiple planets migrating to stable orbits at wide separations.

In Figure \ref{fig:2planet_m}, showing planet masses, the outliers consist of a small cluster of planets with unusually low outer-to-inner mass ratios, potentially indicative of dynamical mixing and a more complex dynamical evolution. (Mass ratios $>$1 are much more common and are predicted by formation models \citep{Dai2020,Weiss23}.) A large minority of systems do have ``inverted'' mass ratios, with the outer planet being lower mass, but most of these are only mildly inverted, with $0.4<M_{\rm outer}/M_{\rm inner}<1$. The outlier systems, however, form a small cluster with $M_{\rm outer}/M_{\rm inner}<1/7$ (the exact value being a break point that appears when also considering higher-multiplicity systems), with the exception of KOI-1783, which is characterized by the rare condition of having a Jupiter interior to a Neptune. Both of these conditions (an extreme mass ratio or a Jupiter interior to a non-Jupiter) we group together as a category of ``strongly-inverted mass ratios.''

We do not classify these strongly-inverted mass ratios as a distinct category of architectures, as they appear too similar to, variously, the ``peas-in-a-pod systems'' and ``warm Jupiter systems'' described in Section \ref{sec:inner}. However, we do consider them as an uncommon and interesting dynamical feature that may indicate significant differences in formation histories. The two-planet systems that fall into this group are highlighted in Figure \ref{fig:2planet_m}, and we discuss strongly-inverted mass ratios further in Section \ref{sec:mixed}.

\section{Higher-Multiplicity Systems}
\label{sec:3plus}

\subsection{Census}
\label{sec:3plusclass}

Prior analyses of the so-called ``peas-in-a-pod'' systems (which comprise the large majority of multiplanet systems) have varied in their coverage of the population, with some papers singling out systems with four or more planets as a distinct population \citep[e.g.][]{Weiss18,Weiss23}, while others group systems of three or more planets as a single population \citep[e.g.][]{GF20}. In our analysis, in our analysis, we group together all systems with \textit{three} or more planets to improve the statistics and draw out patterns. (The number of \four systems in our sample is 112, while the number of \three systems is nearly triple this figure at 314.)

\begin{figure}
    \includegraphics[width=0.99\textwidth]{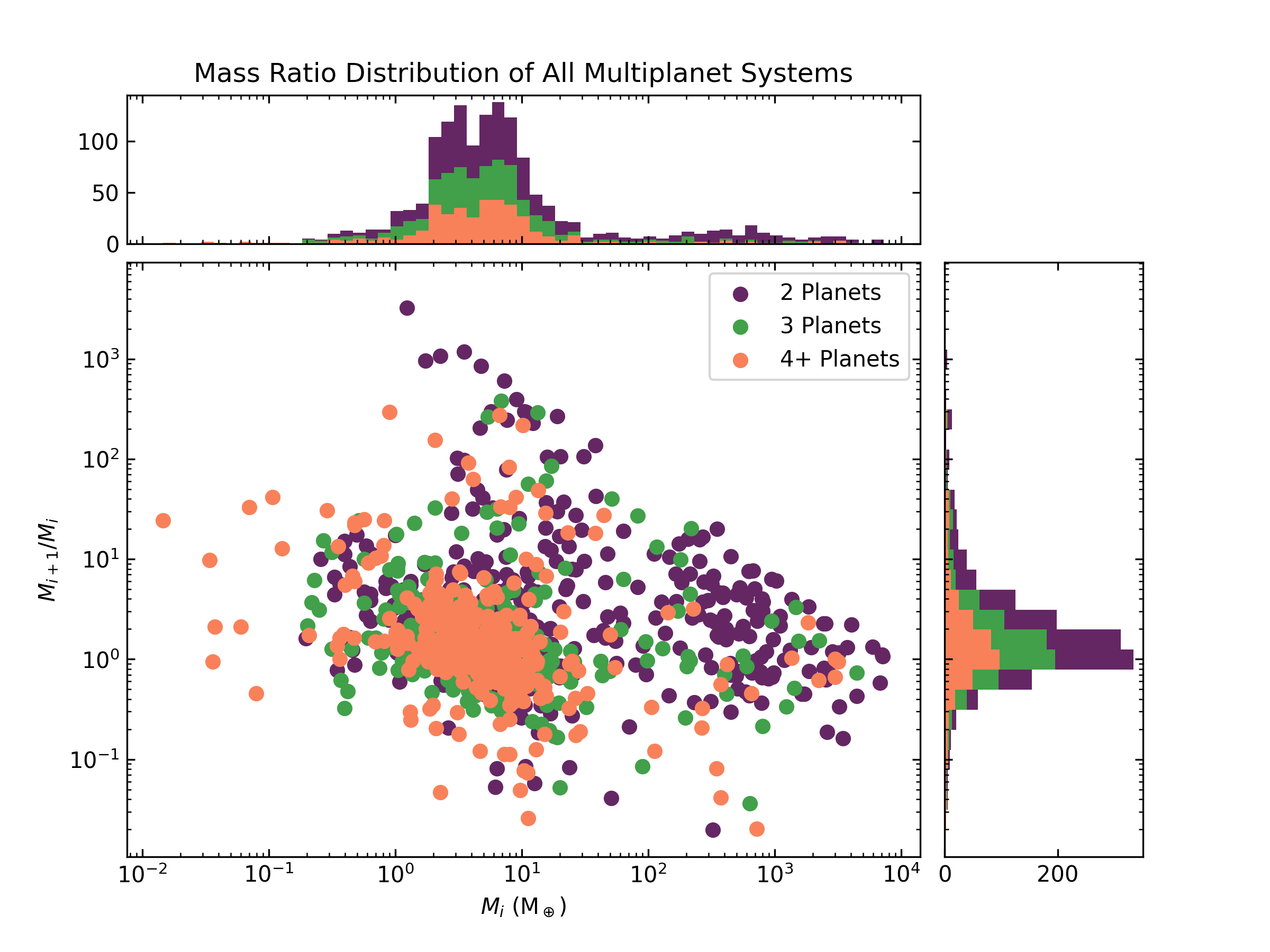}
    \caption{Scatter plot of mass ratios versus inner planet masses for adjacent pairs of planets in 2-planet (purple), 3-planet (green), and higher-multiplicity systems (orange). Each dot represents one pair of planets, for $N-1$ dots per system. The 2-planet systems include more systems with large masses, as evident from the abundance of purple points at high $M_i$ values.}
    \label{fig:234planet_m}
\end{figure}

\begin{figure}
    \includegraphics[width=0.99\textwidth]{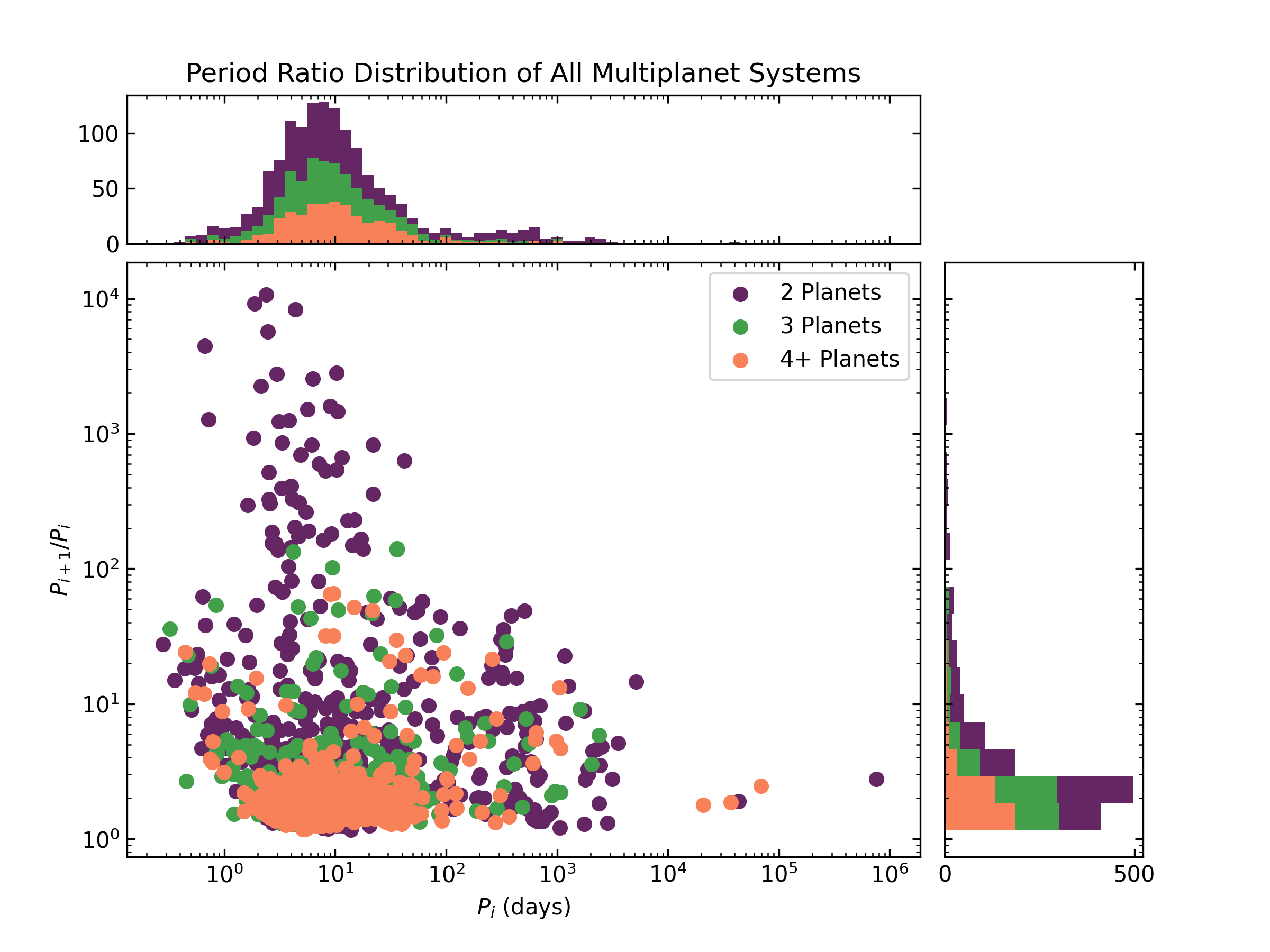}
    \caption{Same as Figure \ref{fig:234planet_m} for period ratios versus inner planet periods. The 2-planet systems show more systems with large period ratios, as evident from the abundance of purple points near the top of the plot. Meanwhile, the higher-multiplicity systems show more similarity to one another in both plots.}
    \label{fig:234planet_p}
\end{figure}

To ensure that this category is statistically meaningful, we repeat our analysis from Figures \ref{fig:2planet_m} and \ref{fig:2planet_p} in Figures \ref{fig:234planet_m} and \ref{fig:234planet_p} for systems with different numbers of planets. These figures show the pairwise period ratios and mass ratios for systems with two planets (purple), three planets (green), and four or more planets (orange, that is, showing all adjacent planet pairs in each system). While this analysis is qualitative, we do indeed see a much greater difference between two and three planets than we do between three and four, as the two-planet systems show systematically more high-mass planets and more planet pairs with large period ratios, while the populations of higher-multiplicity systems are fairly similar to each other. Thus, we can safely group systems with \three planets together.

Planetary systems with \three planets are more difficult to classify than two-planet systems, both because the larger number of planets allows for more dimensions of variation, and because we have grouped together systems with a range of multiplicities (the highest multiplicities being Kepler-90 with eight planets, and TRAPPIST-1 with seven).

There are several ways we could extend the classification scheme in Section \ref{sec:2pl} to apply to higher-multiplicity systems, such as dividing them according to average mass ratio and period ratio, or by the dispersion in mass and period ratios, or dividing them according to the total number of Jupiters in each system. However, any such scheme should take into account the patterns among multiplanet systems that are already known.

\cite{Weiss23} defined the category of ``compact multiplanet systems'' as any system with multiple planets with radii of 0.5-4.0 $R_\oplus$ and periods of one day to one year. They only studied a subset of these systems -- those with four or more qualifying planets -- and they also imposed an additional limitation of tightly-packed planets with $P_{i+1}/P_i<3.5$, primarily to suppress the selection effects from the low transit probabilities of longer-period planets.

This definition is roughly equivalent to similar established categories of multiplanet systems: variously named the ``compact Kepler-type'' systems or the ``peas-in-a-pod'' systems. The large majority of \three systems -- about three quarters of them -- are of this type, with multiple planets, usually of similar size and log-uniform period spacing that are dynamically packed in orbits close to their host star, the prototype for which is Kepler-11 \citep{Kepler-11}.

While the definitions used in different analyses can vary, the criteria are almost always narrower than our scheme in Section \ref{sec:2pl} would suggest. Our framework would suggest setting aside systems with all small (non-Jupiter) planets, even if they aren't particularly uniform, or with all similarly-sized planets, even if they aren't small. On the other hand, our framework does not necessarily require compactness when applied to our entire dataset, as it would suggest grouping any systems with planets that are regularly spaced in log-period, even if they aren't particularly close to their star.

In all of these cases, the distinction is plausibly dynamically important, and in light of this, we first consider a three-by-three grid of planet classes. The three classes of two-planet systems from Section \ref{sec:2pl} can be generalized as systems with only small planets, those with only large planets (Jupiters), and those with a mix of large and small planets. Similarly, we can consider three classes with regard to orbital spacing: systems with all tightly-packed planets reminiscent of the peas-in-a-pod systems, systems with all widely-spaced planets, and systems with irregularly spaced planets with some large and some small period ratios. For the definition of widely-spaced planets, we use a more lenient criterion than \cite{Weiss23} of $P_{i+1}/P_i<5$. We find this to be a meaningful limit in our definition of inner and outer planets as described in Section \ref{sec:io}, and it appears to be a natural break point in the distribution of period ratios for the population as a whole.

The number of planetary systems in each of these nine classes is shown in Table \ref{tab:multis}. We note that the large majority (72\%) of multiplanet systems fall into the category of small, closely-spaced planets, which lines up very well with the traditional ``compact multis.'' However, the situation for the other classes is more complicated.

\begin{table}[htb]
    \centering
    \begin{tabular}{l | r | r | r}
    \hline
        & No Jupiters &  All Jupiters & Mixed Planets \\
    \hline
    All Closely-Spaced & 226 & 3 & 11 \\
    All Widely-Spaced  &   0 & 3 &  5 \\
    Irregularly-Spaced &  33 & 7 & 26 \\
    \hline
    \end{tabular}
    \caption{Preliminary classification of \three planetary systems in a 3-by-3 grid based on the classes defined in Section \ref{sec:2pl}. The ``closely-spaced no-Jupiters'' systems correspond to the standard definition of ``peas-in-a-pod'' or ``compact Kepler-type'' systems, while all others are more irregular in planet size and/or spacing.}
    \label{tab:multis}
\end{table}

First, systems with all widely-spaced planets are so rare that it is not clear if it makes sense for them to have their own categories, especially given that a stability analysis would likely consider such gaps to be ``dynamically detached'' whether they are uniformly spaced or not. From this standpoint, it seems reasonable to fold them into the corresponding ``irregularly-spaced'' categories.

The second problem derives from observational biases. Further observations on a compact Kepler-type system can easily turn a closely-spaced system into an irregularly-spaced system and a no-Jupiter system into a mixed system through the discovery of a distant Jupiter companion. Six systems that would otherwise fall into the ``closely-spaced, no-Jupiters'' category have such a companion (see Figure \ref{fig:peapod}), and it may prove to be a common occurrence with future observations. Thus, the categories in this grid are not fixed, and they are particularly vulnerable to being disrupted by the discovery of outer planets similar to the gas giants of our Solar system. (The opposite case, with previously-undiscovered being found in apparent gaps, is addressed in Section \ref{sec:inner}.)

\subsection{Distinguishing Inner and Outer Planets}
\label{sec:io}

Our own Solar System is readily divided into inner and outer planets between Mars and Jupiter. The subset of exoplanetary systems in our dataset that have planets discovered at a wide range of separations, particularly those that extend past the ice line, show a similar pattern, with one or more giant ``outer'' planets separated by a wide gap from usually smaller interior planets -- with this gap usually being near, but not always at the ice line.

This pattern suggests a classification system that begins by dividing the exoplanet population into inner planets and outer planets, similar to our own Solar System. The inner planets are not necessarily small; there are all-Jupiter systems that have a clear inner-outer divide in their detected planets, most notably HD 27894, which has three giant planets with periods of 18, 36, and 5000 days \citep{Feng22} However, the outer planets \textit{will} be large (we provisionally require the ``first outer planet'' to be a Jupiter) and will be separated from the inner planets by a wide period gap, which is also near the ice line (as opposed to a gap between a warm Jupiter and a hot inner planet).

The exact criteria for defining inner and outer planets are not clear for several reasons. There is the small sample size (only 55 \three systems in our dataset have any detected Jupiters); the degree of variation between systems, where any specific definition can be expected to produce multiple edge cases; and because of the variation in the location of the ice line with stellar spectral type.

An additional complication comes from systems like 55 Cnc, which has four giant planets, the innermost of which has a period of only 15 days, separated by a wide gap from an inner ultra-short period planet \citep{55CncRef}. To avoid a 15-day period becoming an ``outer'' planet, this suggests that a minimum period of the ``outer'' planets would be appropriate, with those with shorter periods being considered ``warm Jupiter'' inner planets, as discussed in Section \ref{sec:inner}. We also need to account for all-Jupiter systems at wide separations that do not have large gaps between their planets, such as HR 8799. In this case, the planets could be counted as outer planets if their periods are all longer than a similar ``minimum outer planet period'' criterion.

When looking at all \three systems with detected Jupiters\footnote{While not directly relevant to this section, we include a figure depicting this population in Section \ref{sec:discuss} as Figure \ref{fig:method}.}, we notice a distinct change in behavior at a period of 100-200 days. Jupiters with periods of $<$100 days are likely to have nearby, low-mass inner companions, where Jupiters with periods of $>$200 days almost never do. This suggests a transition in architectures across the population, with long-period Jupiters showing the gap structure, while short-period Jupiters do not -- and thus suggesting a minimum outer planet period of $\sim$100 days, although again, there is some ambiguity.

There is of course the possibility that additional undetected planets exist in those gaps. However, the evidence of a confirmed gap in our own Solar System along with the statistical evidence presented by \cite{Millholland22} for an outer limit to peas-in-a-pod systems suggest this is likely not the case for most systems. Additionally, the typical period ratio of these inner-outer gaps is much wider than the Mars-Jupiter gap. In fact, the median period ratio under our adopted definition is 32 (for Kepler-65). As this is greater than the square of our minimum gap size, it is too large a gap to be filled by only a single additional planet.

Alternatively, it may be preferable to define inner and outer planets in terms of instellation, preferably at the ice line, given that planet formation processes, especially for giant planets are different on either side of the ice line. However, instellation is an imperfect metric as much as period is because of planetary migration, which can bring giant planets that formed beyond the ice line closer to their host star. Indeed, even Jupiter is believe to have migrated to well inside the ice line temporarily during our Solar System's formation \citep{Chamtela20}. Moreover, in testing, we determined that defining inner and outer planets in terms of instellation instead of period has only a small effect on our classification -- no greater than the difference between a cutoff of 100 days and 200 days. Therefore, for the purpose of this paper, we retain the classification in terms of period.

For the size of the gap in period required to define outer planets, both our own Solar System and the theory of orbital dynamics provide insights. For example, the Jupiter-Mars period ratio is 6.3, so a larger cutoff would fail to capture the dynamics of our Solar System. Meanwhile, the inner edge of the asteroid belt occurs at roughly the 4:1 mean motion resonance with Jupiter \citep{FM94}, indicating some degree of dynamical dominance of Jupiter beyond this distance (that is, with smaller period ratios). While both of these features could easily be accidents of our Solar System's formation, they are indicative of a potential shift in dynamics in the neighborhood of period ratios of $\sim$4-6.

The dynamical stability of planets in binary star systems provides another data point for this question. \cite{HW99} computed numerical stability criteria for orbits of planets in binary star systems in a range of configurations. For high mass ratios and circular orbits (the closest analog to a Jupiter-type planet), they found a critical period ratio of 3.3 for an S-type orbit. Subsequent work \citep{David2003} found a critical period ratio of $\sim$3.7 for Jovian mass companions. Note that these results should be treated with some caution, as they depend on the eccentricity of the planet as well as the age over which stability is enforced. Moreover, these systems can be highly chaotic, so that stability criteria can vary sensitively with system properties. Nonetheless, these results provide good working estimates for the range of period ratios where the gravity of the giant planet is likely to be a significant factor.

The exact choice of period ratio to define the inner-outer divide turns out to be unimportant in the current data set. In moving the cutoff from 4 to 6, only one candidate system for having such a divide is newly excluded: Kepler-46. However, as Kepler-46 would have a minimum outer planet period of only 34 days, it was not in serious consideration to begin with. For definiteness, we adopt a fiducial cutoff of 5 in this paper, as this appears to be a natural break point for gaps in other parts of planetary systems (see Section \ref{sec:inner}), but a different value may prove to be more appropriate once a larger sample of planets has been discovered. For consistency, we also apply this criterion to our definition of ``gapped systems'' in Section \ref{sec:inner}.

In contrast, the cutoff for the minimum outer planet period is more difficult to determine, as several systems fall in or near the approximate cutoff range of 100-200 days, and several others could be plausibly interpreted to define more than one planet as the ``first'' outer planet. Systems where the candidate first outer planet is near this limit need to be analyzed individually in order to determine their status and to adopt a fiducial limiting period for the population:

\begin{itemize}

%\item HD 10180 is the only system that challenges this concept of an inner-outer divide as a whole, as it hosts six planets that straddle the ice line, including a Jupiter at a period of 2205 days, while not having any unambiguous large gaps. (The largest period ratio in the system is 4.9, and the ratio between the sole Jupiter and the next inner planet is only 3.6.) We discuss the unusual features of this system further in Section \ref{sec:outliers}, but it is a large enough outlier that we do not consider it here with respect to out inner-outer definition.

\item Kepler-148 has two low-mass planets with periods of 1.7 and 4.2 days and an apparent warm Jupiter (Kepler-148d) with a period of 52 days \citep{Rowe14}. This is contrasted from other warm Jupiter systems as described in Section \ref{sec:inner}, which either have no gaps among their inner planets, or have a gap with a single planet on the inside. If Kepler-148d were analyzed as an outer planet, the system would better fit the profile of a peas-in-a-pod system with an outer Jovian companion, aside from its unusually short period. However, ignoring the specific multiplicity on each side of the gap, Kepler-148's period distribution is quite typical of gapped peas-in-a-pod systems (as described in Section \ref{sec:inner}), so its apparent unusual status may simply be an artifact of small sample size. Thus, we accept Kepler-148d as an inner planet.

\item HD 33142 has three giant planets with periods of 90, 330, and 810 days. With the largest period ratio between them being 3.7, our scheme would classify them either as all inner planets (in which case 810 days would be unusually long) or all outer planets (in which case 90 days would be unusually short). HD 33142 is a retired A star \citep{Johnson11}, so all three of its planets are well inside the ice line and likely would have been so throughout their lifetimes, so it would appear more plausible to define them as inner planets.

\item HD 141399 has four giant planets with periods of 94, 202, 1070, and 5000 days \citep{HD141Ref}. With its innermost planet near 100 days, it is a candidate to be considered ``all outer planets.'' However, if it were analyzed in this way, it would have an awkward period distribution that straddles the ice line. Instead, the period ratio of 5.3 between the second and third planets is closer to the ice line and appears to be a more appropriate place to draw the inner-outer divide.

\item Similar arguments apply to all of the systems for which the candidate minimum outer planet period is $\lesssim$130 days. HD 141399 is similar in architecture to HD 142, and Kepler-148 is similar to Kepler-25 and Kepler-603.

\end{itemize}

For comparison, of the Jupiter-sized planets with nearby inner low-mass companions (that is, those that do \textit{not} show an inner-outer divide in terms of period ratio), the longest period example is Kepler-289c at 126 days, with two exceptions. Those exceptions are Kepler-90 and HD 10180, both of which we consider to be outliers from the exoplanet population as a whole. We discuss the unusual features of these two systems in Section \ref{sec:outliers}. However, in the current context, we exclude them from our analysis so that all systems with longer-period Jupiters $>$130 days can be cleanly divided into inner and outer planets. For the purpose of this paper, we adopt 130 days as the minimum outer planet period, and by this standard, the shortest-period observed outer planet is HD 14810c, at 148 days.

To be sure, this definition of the inner-outer divide is both arbitrary and over-fit to our small sample. However, our 130 day limit is only a fiducial value chosen to illustrate our classification scheme. A different limit of 80 days or 200 days could be chosen, or a similar limit could be defined in terms of instellation, and it would result in only a slightly different list of inclusion and exclusions in either direction. None of these choices would dispute the \textit{existence} of an inner-outer divide with approximately these parameters, despite the fuzzy edges of the categories. With a larger sample of future exoplanet discoveries, a more robust and physically motivated definition of inner and outer planets could be formulated, and it could determine whether or not there exists a coherent population of ``transitional'' systems.

\subsection{Inner Planets}
\label{sec:inner}

Independent of the exact cutoff between inner and outer planets, nearly all known inner planets have periods $<$1000 days, and the vast majority have periods $<$100 days. This finding means that we have a far larger population of observed inner planets than outer planets, allowing us to make a more thorough analysis.

A full analysis of inner system architectures of the kind we do in Section \ref{sec:3plusclass} requires that we look at systems with at least three \textit{inner} planets, as opposed to three in total. Of the 314 known systems with at least three total planets, 289 (92\%) meet this criterion. By our definition, only one of these (HD 33142) would have an all-Jupiter inner system, leaving only one system in the middle column of Table \ref{tab:multis} and providing good incentive to fold it into the ``mixed planets'' column as a single category. (However, a different definition of the inner-outer divide could change this).

Within this population, we find 23 systems (8\%) with all large planets or a mix of large and small inner planets, while the rest have only small inner planets. Additionally, in both subtypes, we still find some systems with large gaps among their inner planets. This suggests a reduced form of our previous table \textit{restricted to inner planets}, as in the top half of Table \ref{tab:all_inner}.

\begin{table}[htb]
    \centering
    \begin{tabular}{l | r | r}
    \hline
        & Peas-in-a-Pod Systems & Warm-Jupiter Systems \\
    \hline
    Total with 3+ inner planets & 266 &  23 \\
    \hline
    Closely-spaced systems & 232 &  14 \\
    \hline
    Gapped systems &  34 &  9 \\
    \hspace{0.15in} Inner-gap systems      &  18 &   6 \\
    \hspace{0.15in} Middle-gap systems     &   2 &   0 \\
    \hspace{0.15in} Outer-gap systems      &  14 &   3 \\
    \hline
    \end{tabular}
    \caption{Our final classification system for inner planetary systems (limited to those with \three inner planets). The number of closely-spaced ``peas-in-a-pod'' (no-Jupiter) systems is greater than in Table \ref{tab:multis} because peas-in-a-pod type systems with outer Jupiter companions are no longer considered irregular (or ``gapped''). The ``closely-spaced'' and ``gapped'' systems form the core of our classification framework for inner systems, which is extended by more detailed analyses, especially by dividing the gapped systems into inner-, middle-, and outer-gap systems. \\ \textbf{Figure references:} \\ The closely-spaced peas-in-a-pod systems are plotted in Figure \ref{fig:peapod}. \\ The gapped peas-in-a-pod systems are plotted in Figure \ref{fig:small_multis}. \\ The warm Jupiter systems of all types are plotted in Figure \ref{fig:warm_jup}.}
    \label{tab:all_inner}
\end{table}

\begin{figure}[!ht]
    \centering
    %\hspace{0.55in}
    \includegraphics[width=0.99\textwidth]{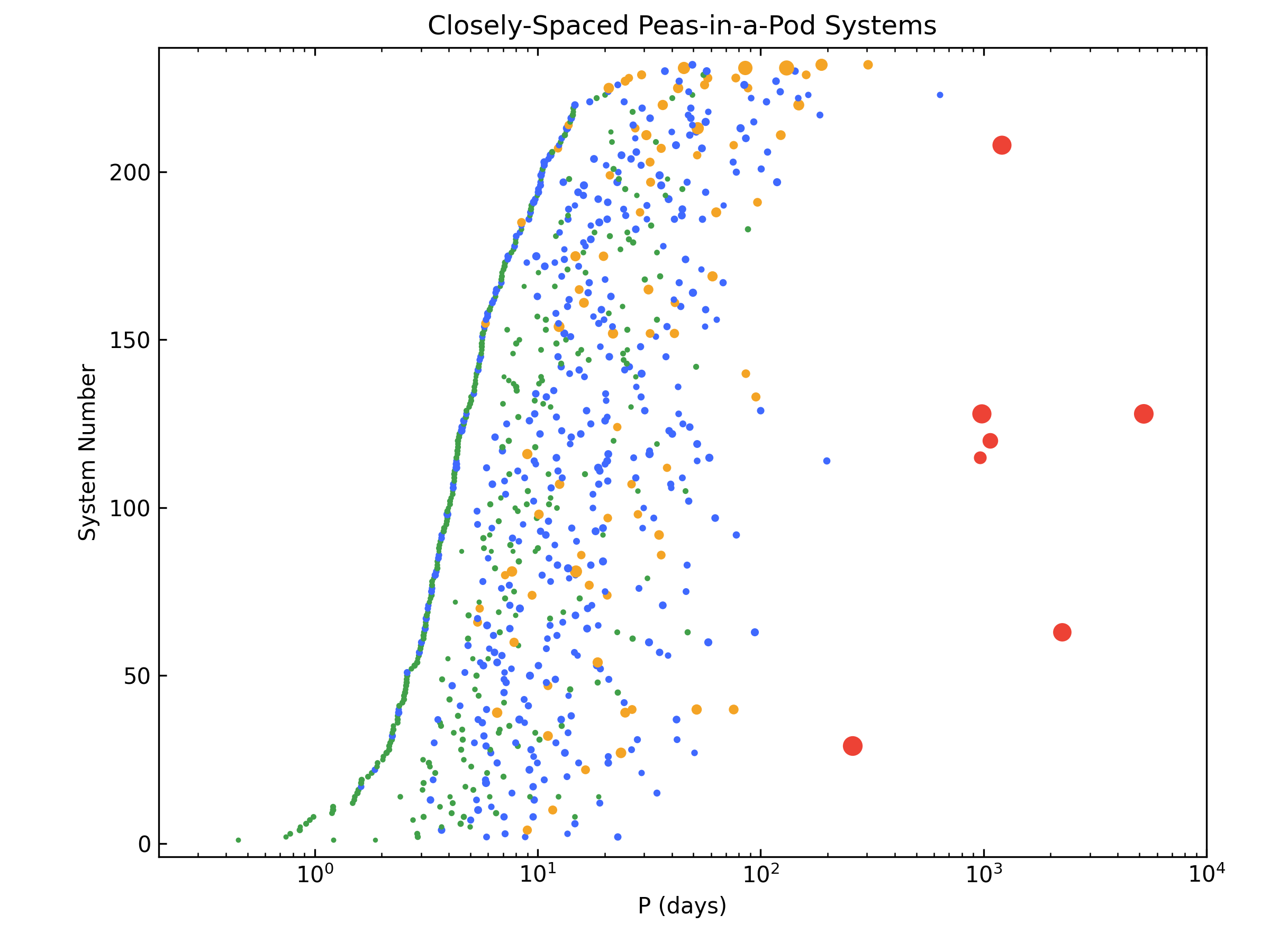}
    \caption{Architectures of all exoplanet systems with \three inner planets that meet the criteria of closely-spaced peas-in-a-pod systems, plotted using the same format as Figure 1. 232 systems are shown (80\% of all systems with \three inner planets), ordered by the period of the innermost planet, bottom-to-top. Six systems have distant Jovian companions, including one with two Jupiters. The shortest-period planet is Kepler-42c. The longest-period non-Jupiter is $\tau$ Cet f.}
    \label{fig:peapod}
\end{figure}

This table forms the basic framework of our classification of inner planetary systems, as discussed in Section \ref{sec:intro}. Systems with no Jupiters among their inner planets we define as ``peas in a pod systems,'' while those with Jupiters we define as ``warm Jupiter systems.'' Likewise, any system with a period ratio $>$5 among its inner planets we define as a ``gapped'' system, while the remainder are ``closely-spaced systems.'' We note that the large majority of \three inner systems (80\%) are closely-spaced peas-in-a-pod systems, consistent with previous analyses (e.g., see the review of \citealt{Weiss23}). A plot of all members of this class of systems is shown in Figure \ref{fig:peapod}.

\begin{figure}[!ht]
    \centering
    %\hspace{0.55in}
    \includegraphics[width=0.99\textwidth]{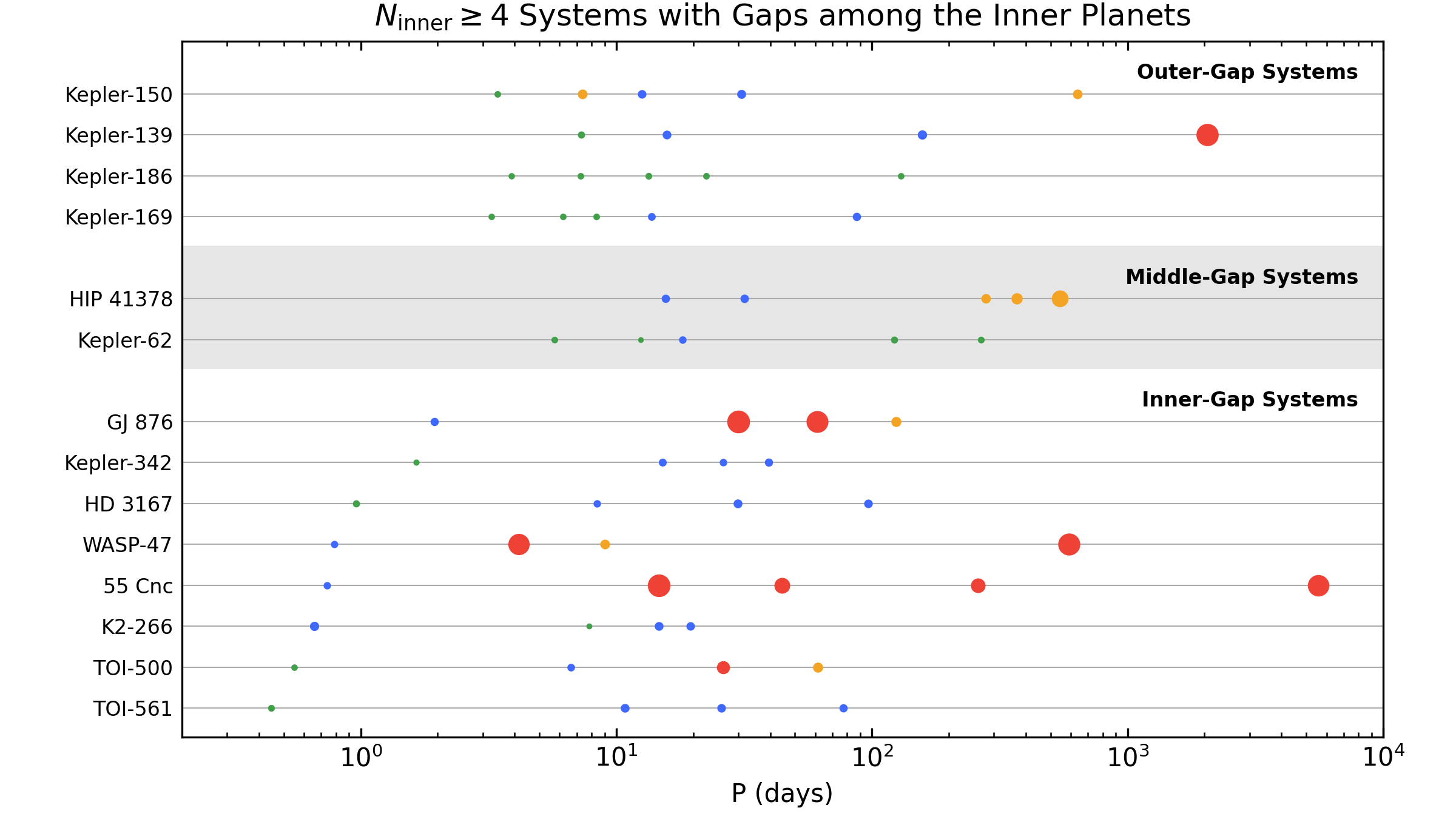}
    \caption{Architectures of the subset of systems with \four inner planets plus a wide gap with a period ratio $>$5, ordered by a location-specific metric of ``gap period.'' This figure shows the relative abundance of inner-gap, middle-gap, and outer-gap systems in a population with commensurable numbers of inner planets.}
    \label{fig:inner4}
\end{figure}

Although the ``gapped'' systems compromise only a small minority (13\%) of the peas-in-a-pod sample, a more detailed picture can be formed by examining this population more closely. In nearly all cases, a closely-spaced inner system becomes a gapped system in one of two ways: the addition of a single hot inner planet, usually small and often a USP (an ``inner-gap system''), or the addition a single cooler exterior planet, usually a sub-Neptune or Neptune (an ``outer-gap system''). ``Middle-gap'' systems are much rarer, although this is complicated by the fact that they require four inner planets to be defined as such. There are 14 systems in our dataset with at least four inner planets and a large period gap (see Figure \ref{fig:inner4}). Of these, eight are inner-gap systems, four are outer-gap systems, and only two (HIP 41378 and Kepler-62) have a large middle gap. (Note that there is statistical observational evidence that the observed outer gaps are indeed ``outer'' -- that is, they should not have additional, undetected exterior planets. Searches for additional planets in compact Kepler-type systems have suggested that our statistics for these systems are largely complete; \citealt{Millholland22}.)

\begin{figure}[!ht]
    \centering
    %\hspace{0.55in}
    \includegraphics[width=0.99\textwidth]{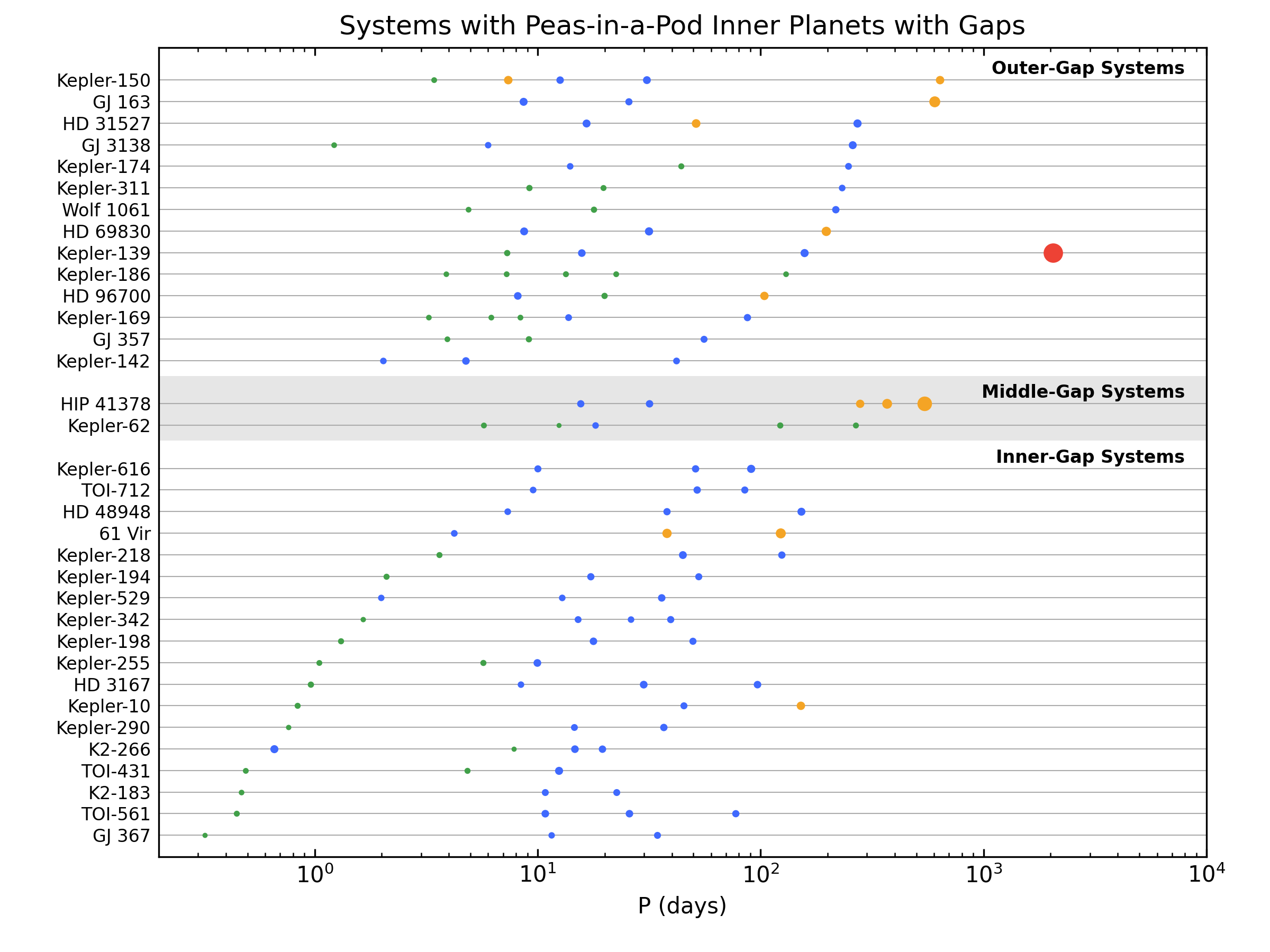}
    \caption{Architectures of gapped peas-in-a-pod systems with \three inner planets, divided by gap location and ordered by a location-specific measure of ``gap period.'' This figure includes all systems in Figure \ref{fig:inner4} that do not have Jupiters among their inner planets, but is extended by systems with only three inner planets (also restricted to non-Jupiters).}
    \label{fig:small_multis}
\end{figure}

\begin{figure}[!ht]
    \centering
    %\hspace{0.55in}
    \includegraphics[width=0.99\textwidth]{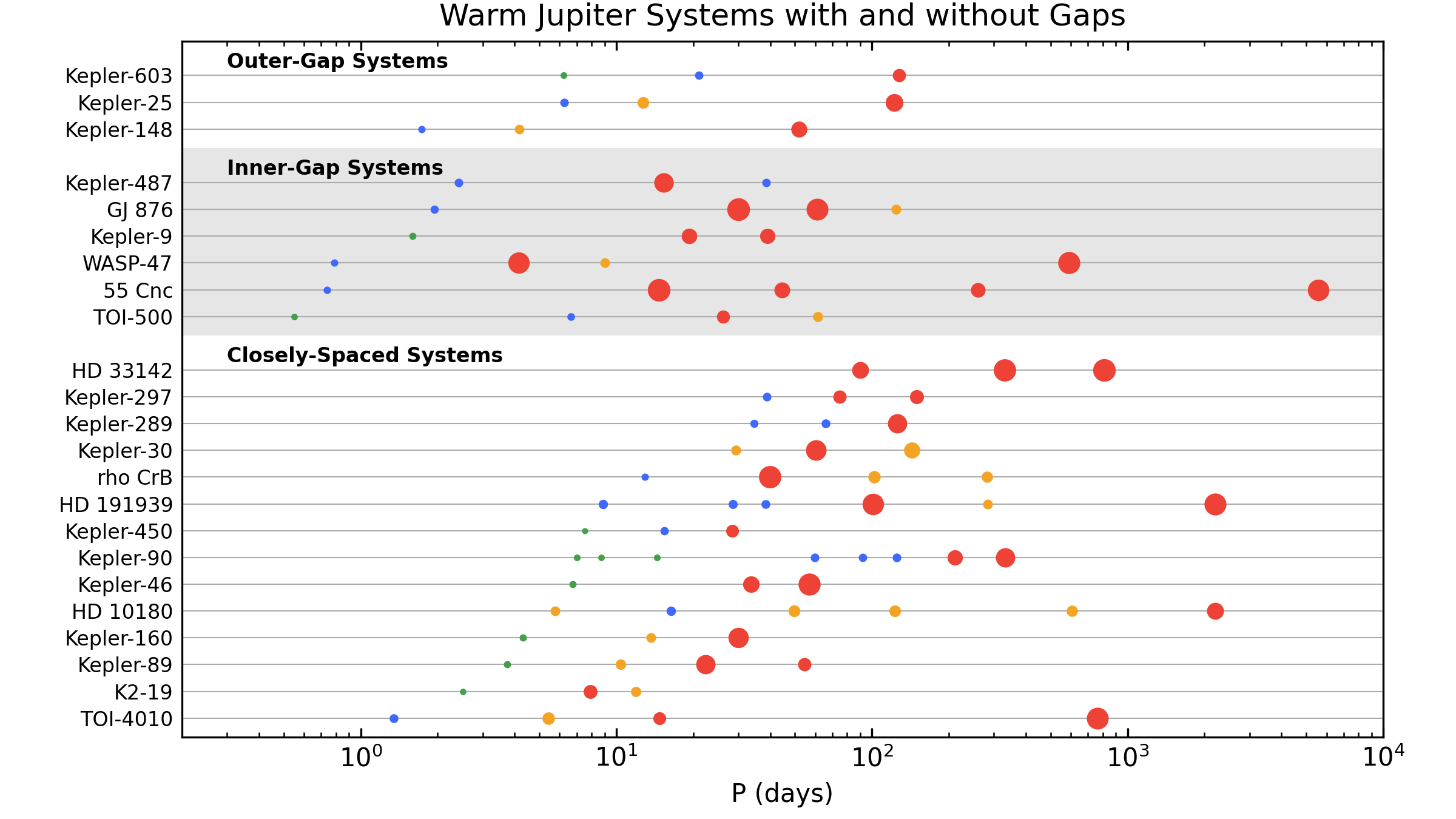}
    \caption{Architectures of warm Jupiter systems (with and without gaps), ordered by period of the innermost planet within each category. This figure includes all systems in Figure \ref{fig:inner4} that have warm Jupiters, but is extended by systems with only three inner planets that also have warm Jupiters. These systems show significantly more variation even within each category than do the peas-in-a-pod systems.}
    \label{fig:warm_jup}
\end{figure}

These distinctions of gap location provide an additional criterion to classify system architectures, instead of simply dividing inner solar systems into closely-spaced and gapped systems. We show this more detailed breakdown in the bottom half of Table \ref{tab:all_inner}. Additionally, we plot the more complete sample of schematics for systems with \three inner planets in two additional figures: Figure \ref{fig:small_multis} for gapped peas-in-a-pod systems and Figure \ref{fig:warm_jup} for warm Jupiter systems (with and without gaps).

Most of the innermost planets in inner-gap systems are USPs or near-USPs with periods of $<$1-2 days, although such cases represent only a small fraction of closely-spaced systems. This trend is consistent with some formation models that predict that USPs will become dynamically detached from the rest of the planets in the system due to gravitational interactions with more distant planets \citep[see esp.][]{Petrovich2019}. This feature contrasts with the outer-gap systems, where one of the planets instead forms farther from the star, or fails to migrate as close to the star, and in extreme cases could be regarded as a low-mass outer planet. This difference suggests that the subdivision of gapped systems by gap location is physically meaningful.

Note that some of the gapped systems could be misclassified if the apparent gaps are actually filled with thus-far-undetected planets, which would reclassify them to the much larger peas-in-pod category. Some population models do predict that such gaps should occur. For example, \cite{He2019} found evidence for clustering of orbital periods in \textit{Kepler} systems. However, without better observations, it is difficult to prove that any individual gap is truly empty. Yet, this does not contradict our results because hypothetical undetected planets are likely to be either small or in highly inclined orbits (e.g., \citealt{Dietrich}), so the systems in question would still differ from the extremely well-ordered peas-in-a-pod systems (albeit in a different way). In the former case, the mass uniformity of the apparently-gapped system is more varied, whereas in the latter case, the system is more dynamically excited than usual. A detailed analysis of the detection limits for possible unseen planets in each system is beyond the scope of this work (see the detailed discussion of \citealt{GF20} regarding this point). However, although some systems might be reclassified, the existence of the categories remains, either by the existence of real gaps in some systems, or by dynamical differences in apparently-gapped systems.

% Chris suggested looking at a mutual Hill sphere criterion, but most of the pairs are separated by many times their mutual Hill radii.

For peas-in-a-pod systems, these categories are clearly delineated, and systems within each class look fairly similar. It's also worth noting that large gaps are uncommon in general (13\%) among the peas-in-a-pod population, compared with a much larger fraction (39\%) of warm Jupiters systems. The exact percentages are subject to large uncertainties given both the small sample size and the observational biases in the detected population, but the difference is large enough to be statistically significant (assuming Poisson noise). %(Adding the Poisson uncertainty in the two cases would yield values of $13\pm2\%$ for peas-in-a-pod systems and $39\pm13\%$ for warm Jupiter systems.)

The difference is even greater for inner-gap systems. If we look solely at inner gaps, we find that they appear in only 7\% of peas-in-a-pod systems, compared with 26\% of warm Jupiter systems. An artifact of our scheme is that in large parts of the parameter space, a Jupiter on the outside of a gap is automatically considered an outer planet and not part of an outer-gap system. Inner-gap systems are not subject to this effect and so provide a better illustration of the differences between warm Jupiter systems and peas-in-a-pod systems.

For other uncommon dynamical features (see Section \ref{sec:other} for details), the disparity is greater still. For example, only eight peas-in-a-pod systems (3\%) exhibit strongly-inverted mass ratios, while eight of the much smaller class of warm Jupiter systems also do (35\%).

We do note one other class of inner systems that mostly does not appear in this \three inner planet sample: the \textit{hot}-Jupiter systems. Hot Jupiters (Jupiters with periods of $<$10 days) are known to be a distinct dynamical class of exoplanets \citep{Wang15}. However, they are usually solitary and so would mostly not appear in the multiplanet system population. Only two \three inner planet systems (K2-19 and WASP-47) contain a hot Jupiter, and of those, K2-19 is an edge case, as K2-19b is very close to our Jupiter-Neptune cutoff at 32.4 $M_\oplus$. We discuss hot Jupiters in more detail in Section \ref{sec:HJ}. However, isolated hot Jupiters without nearby companions could plausibly form a separate, non-overlapping class of systems in our framework.

\subsection{Outer Planets}
\label{sec:outer}

% FRED: the inner-outer divide can't be farther out than the ice line, which is at ~700 days. It can be farther in, to ~100 days, but we have to decide where precisely to draw the line.

In contrast to inner planetary systems, we know much less about outer planets, and what we do know is much more subject to selection effects. With lower transit probabilities, most outer planets are detected by radial velocity, and occasionally by other methods such as microlensing or direct imaging, all of which are much more sensitive to mass. As a result, the vast majority of planets discovered at periods of $\gtrsim$100 days are giants, roughly corresponding to the category of ``cold Jupiters.''

% No inner planets: HR 8799, 47 UMa, HD 184010, HD 11506, HD 37124, HD 142, HD 141399, HD 33142

% 3+ peas in a pod plus outer Jupiter: HD 134606, HD 164922, HD 219134, Kepler-48, Kepler-65, Kepler-167

\begin{figure}[!ht]
    \centering
    %\hspace{0.55in}
    \includegraphics[width=0.99\textwidth]{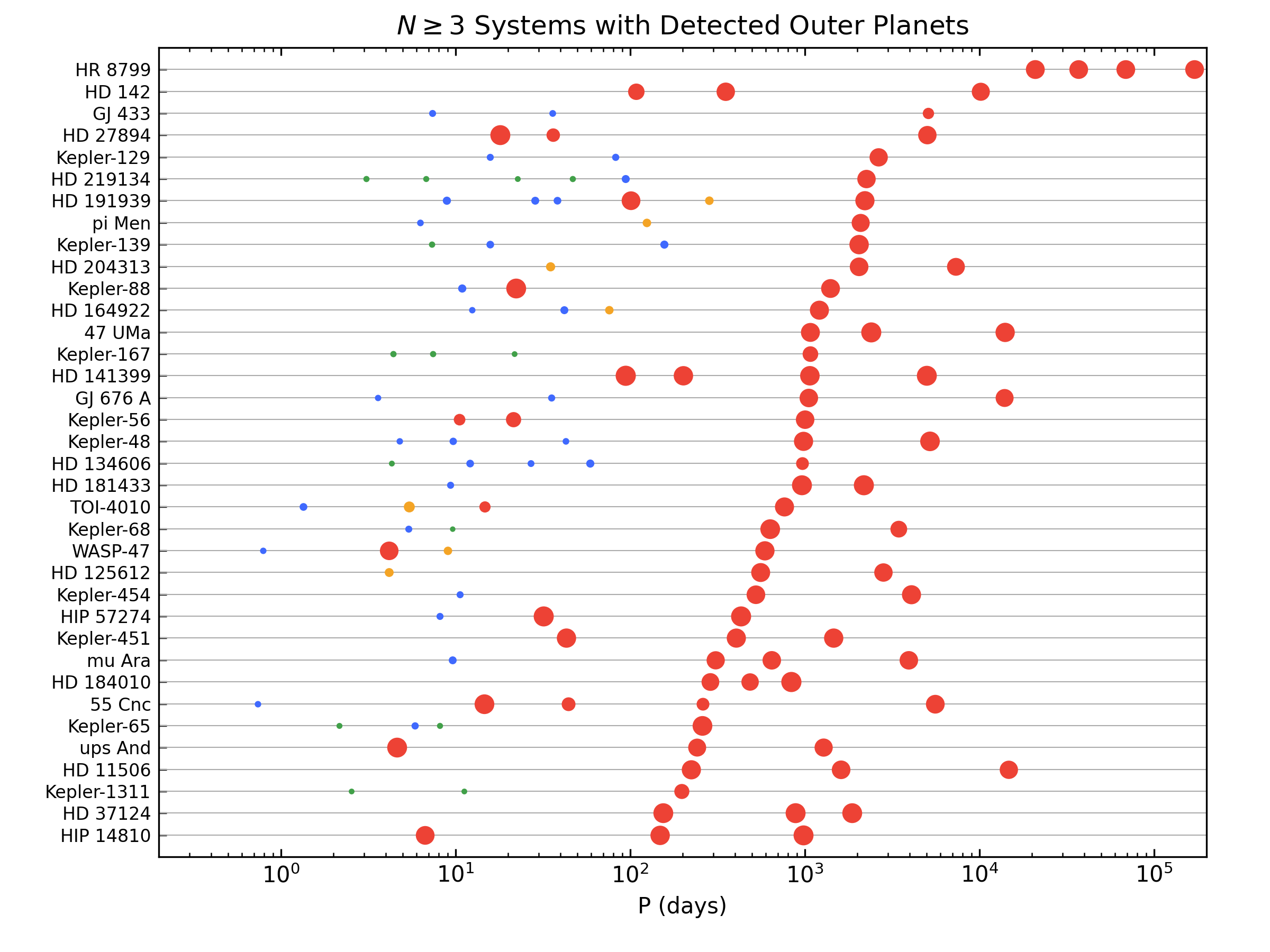}
    \caption{Architectures of \three systems with one or more detected outer planets, ordered by the period of the first outer planet. All known outer planets under our definition are Jupiters, although this is not a definitional requirement. Five systems do not have any detected inner planets. The outer planets are systematically wider-spaced in period ratio than the inner planets.}
    \label{fig:outer}
\end{figure}

Given the smaller sample, in this analysis, we begin by looking at \three systems with one or more planets that meet our definition of ``outer planets.'' There are 36 such systems in our dataset, 31 of which also include inner planets. Roughly half of these inner planets are transiting, so we can hope that this sample is more representative of distant Jupiter companions of the better-characterized inner systems, as opposed to widely-separated giants in general. However, we expect that the selection effects are still substantial. \three systems with detected outer planets are shown in Figure \ref{fig:outer}.

The stricter subset of systems with three or more detected outer planets is too small to draw any clear conclusions. Only six systems meet this criterion, of which only one ($\mu$ Ara) has a detected inner planet. However, one feature of this population is clear, which is that outer planets tend to be more widely spaced than inner planets (including Jovian inner planets). Of the 18 systems in our sample with multiple outer planets, 12 contain a gap with a large period ratio $>$5, and one of them (HD 11506) has two such gaps. (See Table \ref{tab:outer}.) (Our own Solar System is in the minority in this regard, with no period ratios among the outer planets larger than $\sim$3.)

\begin{table}[htb]
    \centering
    \begin{tabular}{l | r}
    \hline
    Total systems with outer planets   &  36 \\
    \hline
    3 or more outer planets            &   6 \\
    \hspace{0.07in} with a large gap   &   4$^a$ \\
    2 outer planets                    &  12 \\
    \hspace{0.07in} with a large gap   &   8 \\
    1 outer planet                     &  18 \\
    \hline
    \end{tabular}
    \caption{Census of outer planets in systems with \three total planets. (All known outer planets under our definition are Jupiters, although this is not a definitional requirement.) \\ \textbf{Figure reference:} all systems with \three outer planets are plotted in Figure \ref{fig:outer}. \\ $^a$HD 11506 has 2 large period gaps.}
    \label{tab:outer}
\end{table}

Some outer-gap peas-in-a-pod systems have planets at sufficiently large distances that they more closely resemble typical outer planets except for not being Jupiters, most notably Kepler-150 and GJ 163. While these could potentially point to another class of outer planets, they are too few in number to indicate a clear pattern to change our current definition, and they could also be removed from such consideration if additional planets are discovered in the gaps.

A deeper understanding of the occurrence rates of outer giant planets \citep[i.e.,][]{Wittenmyer2020} and the covariances of their occurrence with inner companions \citep[i.e.,][]{Zhu2018a, Bryan2019, Bryan2024, VanZandt2024} will improve our ability to resolve these categories in the future.

\section{Additional Dynamical Features}
\label{sec:other}

% From the 09/24/2024 Archive
% Total Hot Jupiters: 688
% Total USPs: 137
% Total USP HJs: 13

In addition to our general classification systems, we note several subpopulations of planets and systems that have unusual or distinctive dynamical features, and which have been studied in past literature. These types of planets can often be identified in lower-multiplicity systems, but we summarize the census of these classes within our better-charaterized sample with \three inner planets in Table \ref{tab:other_types}.

% in this section, we will address additional issues. dynamically defined planet types that may occur in multiple planet types

% additional categories that are important in some way, but not new system categories (except dyn mixed) since they can exist within the previous categories 

\begin{table}[htb]
    \centering
    \begin{tabular}{l | r | r}
    \hline
        & Peas-in-a-Pod Systems & Warm-Jupiter Systems \\
    \hline
    Total with 3+ inner planets        & 266 &  23 \\
    \hline
    With a hot Jupiter                  & N/A &   2 \\
    With an ultra-short period planet   &  16 &   3 \\
    With super-puffs                    &   8 &   1 \\
    With a strongly-inverted mass ratio & 8$^a$ & 8 \\
    \hline
    \end{tabular}
    \caption{Census of unusual dynamical features in systems with three or more \textit{inner} planets. This table demonstrates the relative abundance of these features compared with the other classes of \three inner-planet systems described in Section \ref{sec:inner}. The corresponding figures (Fig. \ref{fig:HJmultis}--\ref{fig:mixed}) include outer planets and lower-multiplicity systems and thus show more systems overall. \\ $^a$Kepler-20 has two planet pairs with strongly-inverted mass ratios.}
    \label{tab:other_types}
\end{table}

\subsection{Hot Jupiters}
\label{sec:HJ}

\begin{figure}[!ht]
    \centering
    %\hspace{0.55in}
    \includegraphics[width=0.99\textwidth]{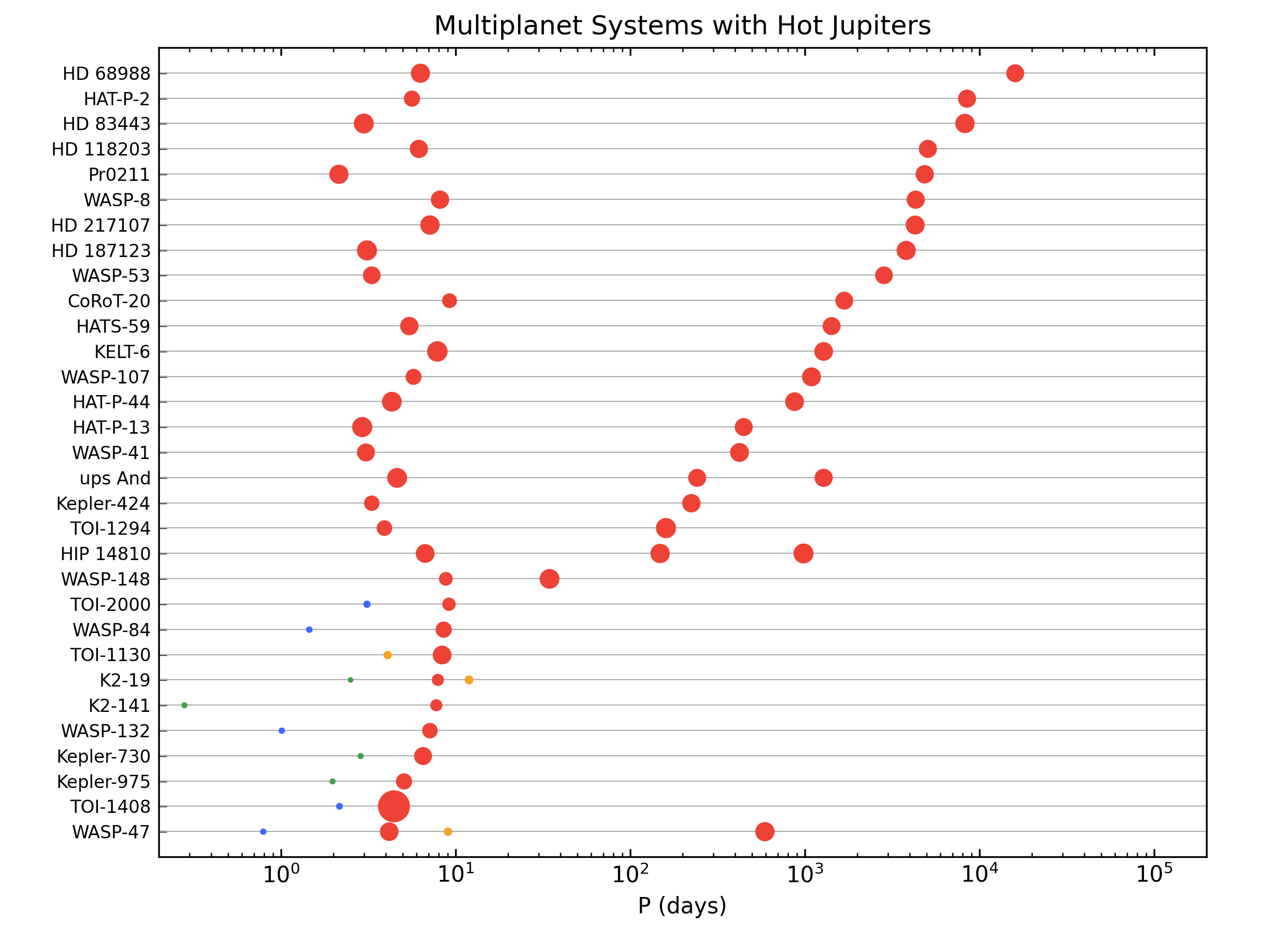}
    \caption{Architectures of multiplanet systems (including 2-planet systems) with hot Jupiters ($P<10$ days), ordered by the period of the second planet in the system to illustrate the distribution of companions. In this sample, this is always the hot Jupiter in systems with nearby smaller companions, and always the distant Jovian companion in systems that have them (except WASP-47). Only WASP-148 has a nearby warm Jupiter companion.}
    \label{fig:HJmultis}
\end{figure}

Hot Jupiters, Jupiter-sized planets with periods $<$10 days, have multiple proposed formation pathways, including \emph{in situ} formation \citep{Batygin2016}, dynamically quiet disk migration \citep{Lin1996}, and high-eccentricity migration \citep{Trilling1998}. While \emph{in situ} formation and disk migration can account for some hot Jupiters, recent evidence suggests that high-eccentricity migration is the dominant process \citep{Zink2023}. This mechanism involves initial dynamical excitation by a companion (planet or star), followed by tidal circularization due to interactions between the planet and its star. %Notably, Lidov–Kozai oscillations, triggered by inclined companions, can induce such migrations, though they likely explain only a small fraction (around 20\%) of hot Jupiter cases. More commonly, coplanar high-eccentricity migration, where even initially coplanar outer companions perturb cold Jupiters into hot orbits, seems to be the prevalent pathway.
The multiplanet hot Jupiter population (including two-planet systems), as shown in Figure \ref{fig:HJmultis}, demonstrates that most (68\%) known multiplanet systems that host hot Jupiters also host cold Jupiters. Since most hot Jupiters are believed to form through high-eccentricity migration, these cold Jupiter companions could potentially serve as the source of the scattering events that initiate this migration process.

However, transit data from Kepler and TESS have identified a small number of systems (10 in our sample\footnote{We note that K2-141, included in this subset, contains a planet that fits the definition of a hot Jupiter by radius, but mass measurements indicate that it may be a much smaller planet. As a result, this system may not be a \textit{bona fide} member of this dynamical class.}) where hot Jupiters coexist with close planetary companions \citep[see the bottom rows of Figure \ref{fig:HJmultis};][]{Becker2015,Morton2016, Canas2019,Huang2020, Hord2022, Sha2023, Maciejewski2023,Bonomo2023, Korth2024}, a geometry inconsistent with the high-eccentricity mechanism that appears to have produced the majority of the sample. These findings imply that less violent processes like \emph{in situ} formation or disk migration could also play significant roles in the formation of hot Jupiters. 
Thus, while high-eccentricity migration appears dominant, the diversity of observed system architectures suggest that multiple formation pathways may contribute to the population of hot Jupiters.
Notably, of this subset, WASP-148 is the only known system to have a hot Jupiter-warm Jupiter pair \citep{Hebrard2020}, i.e. a hot Jupiter with a Jovian companion that is \textit{not} dynamically detached.

Crucially, distant Jovian companions have been discovered for hot Jupiters at roughly the same rate as peas-in-a-pod system, specifically, 21 out of 688 hot Jupiters (including WASP-148) versus 7 out of 232 peas-in-a-pod systems, both very close to 3\%, within the margin of uncertainty. In both cases, the presence of these outer companions has significant impacts on the details of the architectures of the inner systems: distant Jovian companions are thought to be the origin of the scattering events that take hot Jupiters to their final orbits \citep{Zink2023}; for peas-in-a-pod systems, similarly, the distant giant companions are correlated with higher gap complexity in the inner system \citep{HeWeiss}. 

% followup project: check for this list which are feasible origins of the scattering event

\subsection{Ultra-Short Period Planets}

\begin{figure}[!ht]
    \centering
    %\hspace{0.55in}
    \includegraphics[width=0.99\textwidth]{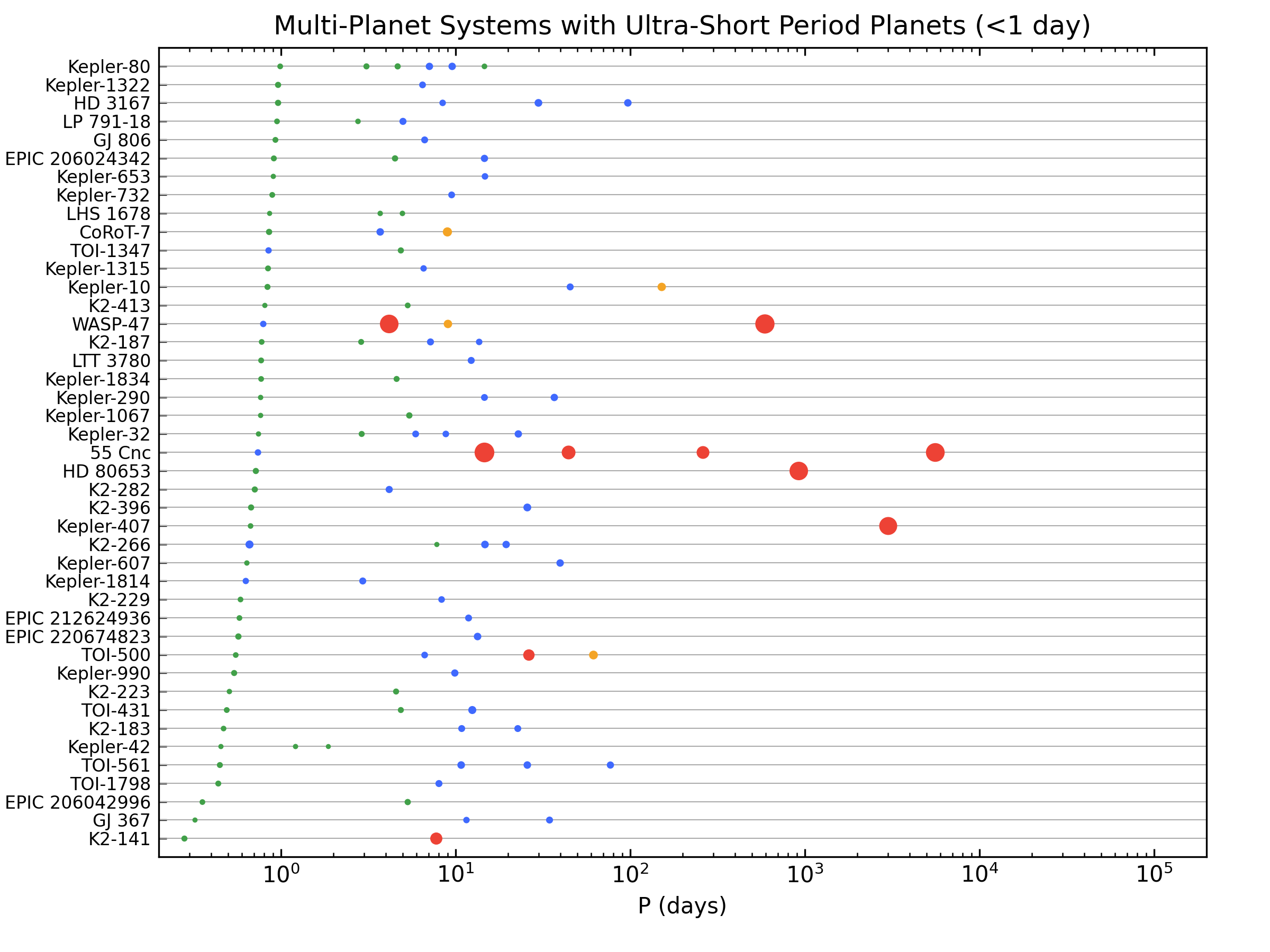}
    \caption{Architectures of multiplanet systems (including 2-planet systems) that include an ultra-short period planet (USP) with $P<1$ day. The USPs have a variable degree of dynamical detachedness from the other (usually peas-in-a-pod) planets in their systems. Kepler-42 is the only system with additional planets within $P<2$ days.}
    \label{fig:usp}
\end{figure}

Ultra-short-period planets (USPs) are a subset of exoplanets characterized by their extremely short orbital periods, defined as being less than one day \citep{Winn18}. The known multiplanet systems with USPs are shown in Figure \ref{fig:usp}. These planets are generally small, with radii $<2 R_{\oplus}$ \citep{Winn18}. Only a small fraction of USPs are ultra-hot Jupiters \citep[13 out of 137 in our dataset; see also][]{Nielsen2020, Yee2023}. USPs as a whole are rare, orbiting roughly 0.5\% of G-dwarfs \citep{SanchisOjeda2014}. 
USPs are often found in multiplanet systems \citep{Adams2021}. In the sample considered in this work, 43 planets have USP orbits and also occur in a multiplanet system (31\% of all USPs compared with 5\% for hot Jupiters and zero ultra-hot Jupiters), a population that is visualized in Figure \ref{fig:usp}. However, when these planets do occur in multiplanet systems, they exhibit distinct dynamical characteristics compared to their neighboring planets. This distinction is evident in their orbital inclinations \citep{Dai2018} and semi-major axis ratios with nearby outer companion planets \citep{Winn18}, suggesting that they are often dynamically decoupled from the other planets in their systems. This decoupling might be a result of unique formation and evolutionary pathways that separate them from more typical planetary bodies within the same system \citep[e.g.][]{Petrovich2019, Pu2019, Millholland2020, Becker2021}. 

The most extreme USPs, particularly those with the shortest orbital periods, may have decaying orbits \citep{Dai2024} or be in the process of disintegration \citep{Rappaport2012, SanchisOjeda2015}. These dynamical states generally require further analysis to confirm, as such inferences rely on accurate determination of the planets' orbital parameters and transit timing and duration variations. 

Interestingly, while the overall population of USPs shares some common characteristics, there is significant diversity within this group. 
The more massive among them, the ultra-hot Jupiters and ultra-hot Neptunes \citep{Nabbie2024}, likely formed through different processes compared to the majority of the sample with $<2 R_{\oplus}$ -- and none of these most massive USP planets are yet observed to occur in multiplanet systems. The smaller USP planets, which do often occur in multiplanet systems, have radii shaped by processes such as photoevaporation \citep{Dai2021}. Only a fraction of these are theorized to be remnant cores of gas giants \citep{Konigl2017, Uzsoy2021}, with the rest likely being born as smaller planets.  

\subsection{Super-Puffs}

\begin{figure}[!ht]
    \centering
    %\hspace{0.55in}
    \includegraphics[width=0.99\textwidth]{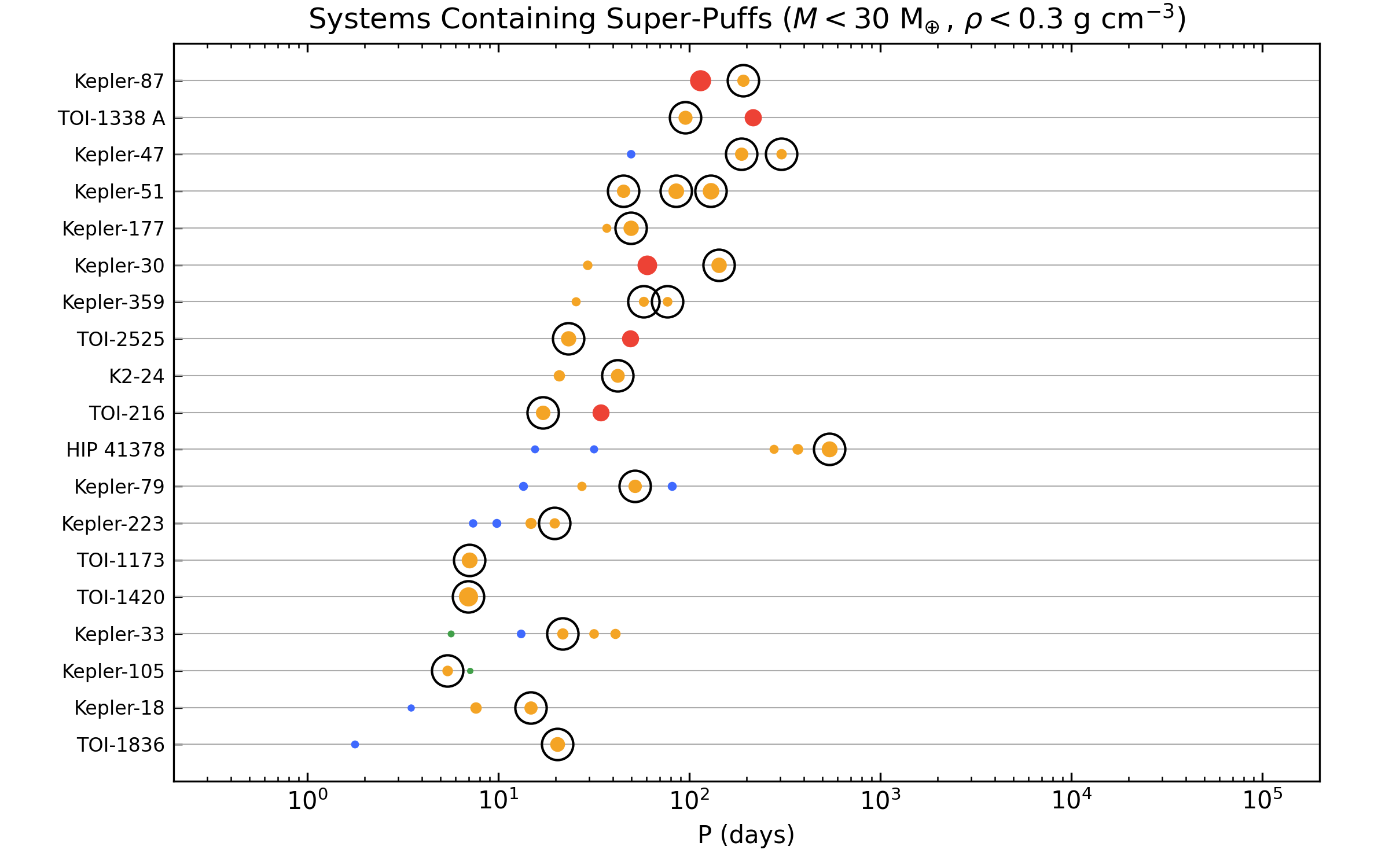}
    \caption{Architectures of planetary systems (of any multiplicity) that contain super-puffs (circled), ordered by period of the innermost planet. We define super-puffs as $<$0.3 g cm$^{-3}$ and $<$30 M$_{\oplus}$. All super-puffs are classified as ``Neptunes'' in our analysis regardless of other properties.}
    \label{fig:puff}
\end{figure}

The population of planets with anomalously low densities and low masses ($<$0.3 g cm$^{-3}$ and $<$30 M$_{\oplus}$ in our analysis) are colloquially called ``super-puffs'' \citep{SuperPuffDef}. 23 planets in our dataset meet these criteria. The known population of super-puffs is shown in Figure \ref{fig:puff}, including single-planet systems, although only two of the 19 systems known to contain super-puffs are single-planet systems. Notably, super-puff planets typically occur in systems with broadly mixed densities, with the exception of the young Kepler-51 system.
Individual super-puff planets in multiplanet systems can be explained by various origin hypotheses depending on their specific parameters. These include young planets with high envelope entropy and hazes (Kepler-51; \citealt{Masuda2014, LibbyRoberts2020}), those with substantial hydrogen and helium atmospheres (Kepler-47; \citealt{Orosz2019}), those inflated by stellar irradiation \citep{Laughlin2011, Thorngren2018}, and others with unexplained origins. Short-period super-puffs that are the innermost planets in their systems, like TOI-1420b \citep{Yoshida2023, Vissapragada2024}, have radii that can be attributed to their high equilibrium temperatures and substantial stellar irradiation. In contrast, long-period super-puffs around old stars in multiplanet systems have more ambiguous origins. For instance, in the HIP 41378 system, the sole super-puff is the outer planet, and planetary rings have been suggested as a possible explanation for its apparent density and low derived possible core mass \citep{Piro2020, Akinsanmi2020, Belkovski2022}.

% WASP-107b was excluded in the 9/24 analysis because of a revised mass estimate.

Due to the wide variety of physical explanations for why a planet might be a super-puff, this sub-population cannot be considered as a coherent group, unlike the other populations described in this work. Instead, each planetary system must be individually analyzed. Depending on the origin of the low-density planet in each system, these systems may also fit into other categories defined in this work.

% these ones are unique because there are sub-populations, with origins that make varying amounts of sense. 

% distinct hypotheses for their densities 
% if evaporation picture is correct, they will only evaporate if mag fields are not too strong, which may supress outflow
% problem for future: mag fields could protect super puffs and so you can use that to constrain their fields 

\subsection{Strongly-Inverted Mass Ratios}
\label{sec:mixed}

\begin{figure}[!ht]
    \centering
    %\hspace{0.55in}
    \includegraphics[width=0.99\textwidth]{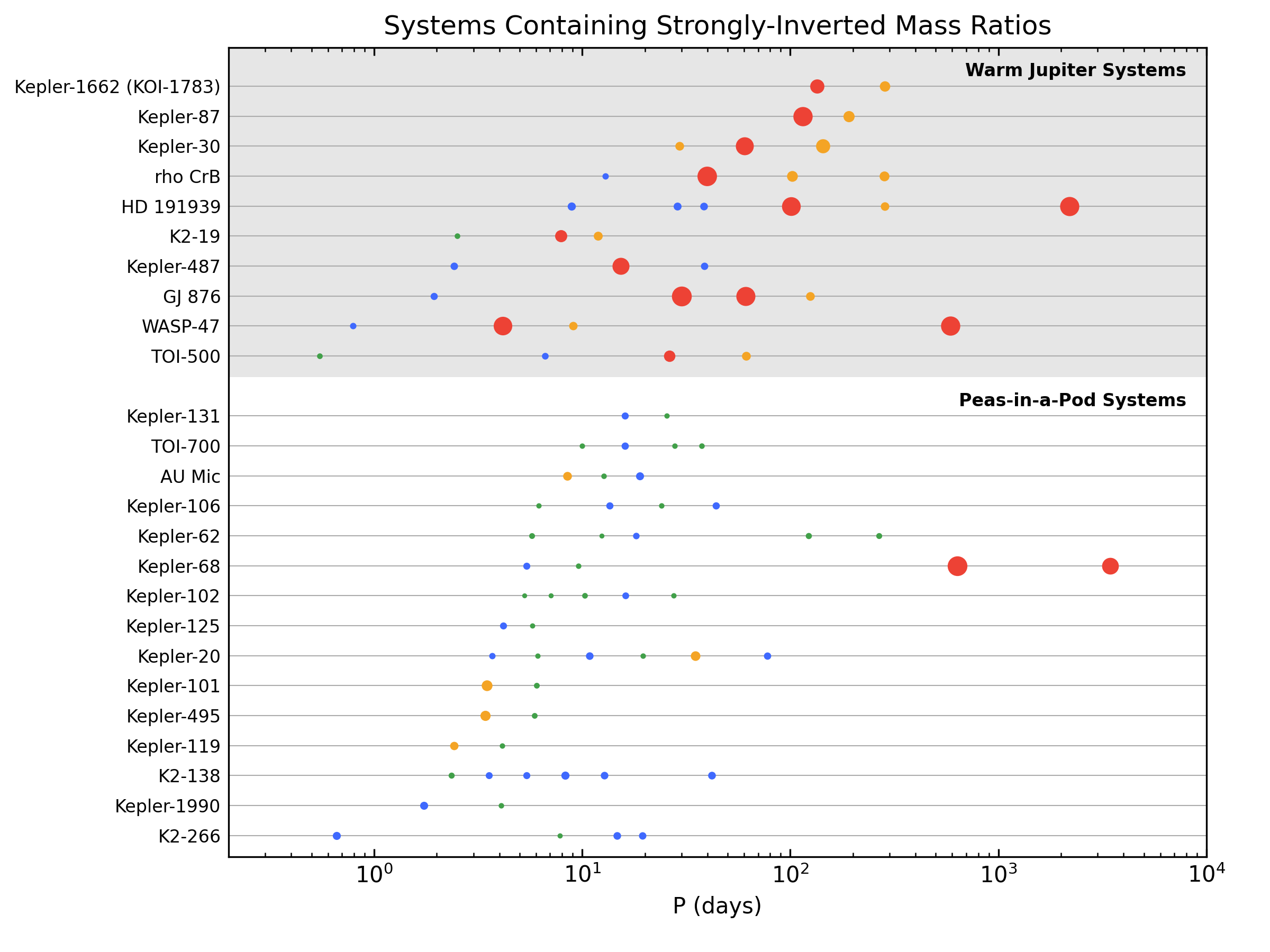}
    \caption{Architectures of multiplanet systems (including 2-planet systems) that have planet pairs with strongly-inverted mass ratios, ordered by period of the innermost planet in each class. We define these systems as warm Jupiter systems with a non-Jupiter exterior to a Jupiter (top) or peas-in-a-pod systems with a pair of adjacent planets with an outer-inner mass ratio of $<1/7$. All such planets under our definition are inner planets because no non-Jovian outer planets appear in our dataset.}
    \label{fig:mixed}
\end{figure}

We define a strongly-inverted mass ratio in two ways depending on the sizes of the planets involved. The first is a Jupiter with an outer companion that is smaller than a Jupiter. 10 systems have planet pairs that fall into this category. For non-Jupiter pairs, we define two adjacent planets to have a strongly inverted mass ratio if $M_{\rm outer}/M_{\rm inner}<1/7$. (This is an empirical result based on looking at the distribution of mass ratios in multiplanet systems.) 15 systems fall into this category, with Kepler-20 having two such pairs. All 25 systems in our dataset with strongly-inverted mass ratios are shown in Figure \ref{fig:mixed}. All such strongly-inverted mass ratio pairs occur among inner planets, as all outer planets in our dataset are Jupiters and therefore would not fit either definition.

We consider strongly-inverted mass ratios to be potential evidence for dynamical scattering events. However, this is not always the case; it is worth noting that our own Solar System itself contains two pairs of planets that meet this definition: Earth-Mars and Saturn-Uranus, neither of which appears to be the result of scattering. The small size of Mars, while difficult to replicate in simulations, is considered to be an artifact of formation, not scattering \citep{Tsiganis15}. And Uranus may have switched places with Neptune during the Solar System's formation, but not Saturn, and formation models predict that planets formed in the outer parts of the disk should be smaller regardless \citep{Tsiganis05}. (This may necessitate an adjustment to our definition in the future, but as current detection methods cannot detect Uranus analogs, we do not need to worry about their impact at this time.)

Strongly-inverted mass ratios nevertheless remain a rarity in our exoplanet dataset, and they appear as a distinctive dynamical cluster. The origin of these strongly-inverted mass ratios is not yet clear, and some of them may be a result of dynamical processes such as planet-planet scattering, as well as the gravitational influence of outer giant planets \citep{Quintana07}. However, the fact that two such pairs appear in our own Solar System suggests that they may instead be much more common and only appear rare due to observational biases. A more complete detection survey of low-mass planets is needed to disentangle these possibilities.

%We consider only adjacent pairs of planets in this way both because the clustering effect appears only in that sub-population and because pairwise adjacency is more dynamically significant. 

\subsection{Outliers}
\label{sec:outliers}

An important feature of the classification system presented in this paper is that it can categorize nearly all known planetary systems that have sufficient data. With rare exceptions, nearly all of them will fall into an appropriate category, although it is not always obvious how to draw the categories themselves to maximize their effective coverage of the dataset. For example, 55 Cnc and WASP-47 showed up again and again looking strange in our attempts at plotting different classes of systems, but when properly divided into inner and outer planets, they appear much more normal. Others, like GJ 411, may appear to be outliers in one population, but look much more typical in a broader population.

Our own Solar System is no exception to this. While we have struggled to find any exoplanets similar to Earth (an Earth-sized planet near 1 AU around a Sun-like star), our Solar System is fairly typical in that it has a ``peas-in-a-pod'' inner system (albeit less uniform than average) separated by a large gap from an outer system of four giant planets (two of which are detectable with current methods).

However, there are a few planetary systems that appear unusual by any analysis, and that fit into this framework poorly or not at all, even on the tail ends, which are summarized in Table \ref{tab:oddballs}. These systems merit further analysis:

\begin{itemize}

\item HD 10180, by our definition, is technically a closely-spaced warm-Jupiter system. However, analyzing it in this way assigns HD 10180h to be by far the longest-period ``warm Jupiter'' in our sample at 2205 days \citep{HD10180ref}. It is also the only system in our dataset where ``inner'' planets unambiguously extend past the ice line. Instead, the HD 10180 system looks much more like a typical peas-in-a-pod system, scaled up to the point where most of its planets are Neptunes. HD 10180 is a remarkably sun-like star, with only a 6\% greater mass and 20\% greater metallicity \citep{Tuomi12}, so this may suggest unusual features of its planet formation environment.

\item HIP 41378 is a system of five planets discovered via transits. It hosts a long-period super-puff (the longest such period currently known), but all other planets in the system have more typical densities \citep{HIP41378}. The system has a large period ratio ($\sim$9) between the second and third planets, unusually separating two sub-Neptunes from three Neptunes, potentially resembling an outer planetary system, scaled down, although this is uncertain. There is a candidate, low-SNR, non-transiting planet in the gap between those two planets \citep{HIP41378}, which has not yet been confirmed. The candidate Planet g, if confirmed, would reclassify the system as a standard closely-spaced peas-in-a-pod system, but even then, the presence of such a long period super-puff would remain anomalous.

%Kepler-62 and HIP 41378 are also in the category of rare middle-gap systems. Kepler-62 is an otherwise-normal peas-in-a-pod system and so can be properly classified by our system, but HIP 41378 is doubly notable for its unusual structure with three Neptunes spaced close together exterior to the gap, one of which is also the longest-period known super-puff at 542 days (and also the outermost known planet in its system).

\item HR 8799 remains the only system with four (or even three) directly-imaged planets. Widely-separated planets may be quite common, but having four of them in near-circular orbits at such large distances (16-72 AU) suggests unusual formation and/or migration processes \citep{HR8799ref}.

\item Kepler-42, while it has only one true ultra-short-period planet (with a period shorter than one day), is the only known system with multiple planets with periods shorter than \textit{two} days, of which it has three \citep{Muirhead12}. Kepler-42 is a mid M-dwarf with a mass of only 0.144 $M_\odot$, but even so, for its planetary system to be so compact is still an outlier in the population compared with e.g. TRAPPIST-1.

% TRAPPIST-1 has the next shortest-period second planet at 2.4 days.

\item Kepler-51 is unremarkable dynamically, but it is unusual because all three of its planets are super-puffs (the most of any known system), and all three of them are also the lowest-density planets in our dataset. Despite the system's young age of 500 Myr, this also suggests a potentially unusual planet formation process, which has not yet been resolved \citep{Masuda2014, LibbyRoberts2020, Lammers2024}.

\item Kepler-90 by our definition is technically a closely-spaced warm-Jupiter system. However, this framework would give it two unusually long-period warm Jupiters, at 210 and 330 days \citep{K90ref}, surpassed only by HD 10180 and the evolved HD 33142 system. It also contains a gap with a period ratio of 4.1, which while neither in the right place nor large enough to divide an inner and outer system by our definition, still looks odd when compared with other warm-Jupiter systems. The fact that Kepler-90 has three closely-spaced Earths interior to this gap and three closely-spaced sub-Neptunes plus two Jupiters exterior to it gives it one of the most unusual dynamical structures in the dataset, although further discoveries of high-multiplicity systems may be needed to determine how anomalous it truly is. (Kepler-90 is also the only confirmed eight-planet system besides our own.)

\item Finally, WASP-148 is the only known hot Jupiter with a nearby Jovian companion (that is, not dynamically detached, \citealt{WASP148ref}). Given the high detection probability of such a companion and the fact that 10 hot Jupiters are known to have smaller nearby companions, this points to an especially rare subtype of system and potential unusual migration processes.

\end{itemize}

While not outliers \textit{per se}, we also note two systems that draw interesting contrasts with two of these outliers. TRAPPIST-1, in addition to being the only confirmed seven-planet system \citep{TRAPPIST}, contrasts with HD 10180 in that it looks remarkably like a ``typical'' peas-in-a-pod system, scaled \textit{down}. Meanwhile, $\tau$ Ceti contrasts with Kepler-42 in that it is the longest-period peas-in-a-pod system currently known, with periods (in the four-planet solution we adopt, \citealt{TauCet}) ranging from 20 days to 636 days, which therefore makes it the most similar known solar system to our own.

\begin{table}[htb]
    \centering
    \begin{tabular}{l | l}
    \multicolumn{2}{c}{The Seven Oddballs} \\
    \hline
    System & Claim to Fame \\
    \hline
    HD 10180 & Inner planets extend past the ice line, resembles an oversized peas-in-a-pod system \\ % closely-spaced warm Jupiter
    HIP 41378 & Longest-period super-puff, unusual dynamical structure \\ % middle-gap peas-in-a-pod / super-puffs
    HR 8799 & Most planets at wide separations \\ % outer system
    Kepler-42 & Most compact multiplanet system \\ % closely-spaced peas-in-a-pod / ultra-short period
    Kepler-51 & Three super-puffs, which are also the three lowest-density known planets \\ % closely-spaced peas-in-a-pod / super-puffs
    Kepler-90 & Only confirmed 8-planet system, unusual dynamical structure \\ % closely-spaced warm Jupiter
    WASP-148 & Hot Jupiter with nearby Jovian companion \\ % hot Jupiter
    \hline
    \end{tabular}
    \caption{Features of the seven planetary systems that either challenge our classification system or that may be easily classified, yet still appear as outliers in the population.}
    \label{tab:oddballs}
\end{table}

% Discussion section? Need a well-considered discussion of possible biases.

% All-Jupiter systems are likely to have good RV data and good limits one smaller planets.

% Michael Meyer: what are the overall occurrence rates for warm Jupiters, and how do they compare?

% Fred: what can we say about possible future discoveries? (e.g. longer baselines)
% Alex: e.g. the Batygin & Morbidelli formation model predicts peas-in-a-pod systems are bimodal
% Fred: other formation models predict it's more favorable for one planet to eat all the mass at longer periods.
% Alex: longer baselines can test formation models, but they could show up better in architecture-based analyses than overall occurrence rates.

\section{Observational Biases and Avenues for Future Expansion}
\label{sec:discuss}

% what about binaries? all single stars have planets (kepler), but not all binaries do... what fraction is that? (we don't know, so this is future work) 

While the framework presented in this paper is very effective at classifying the known population of exoplanet systems, which was used to define it, it is necessarily limited by the kinds of planets we are able to detect, which are extremely non-uniform across the population. It is also limited by the large observational biases within the population of detected planets, which differ between detection methods. At the same time, this framework allows the possibility of additional classes of planets that have not yet been discovered.

\begin{figure}[!ht]
    \centering
    %\hspace{0.55in}
    \includegraphics[width=0.99\textwidth]{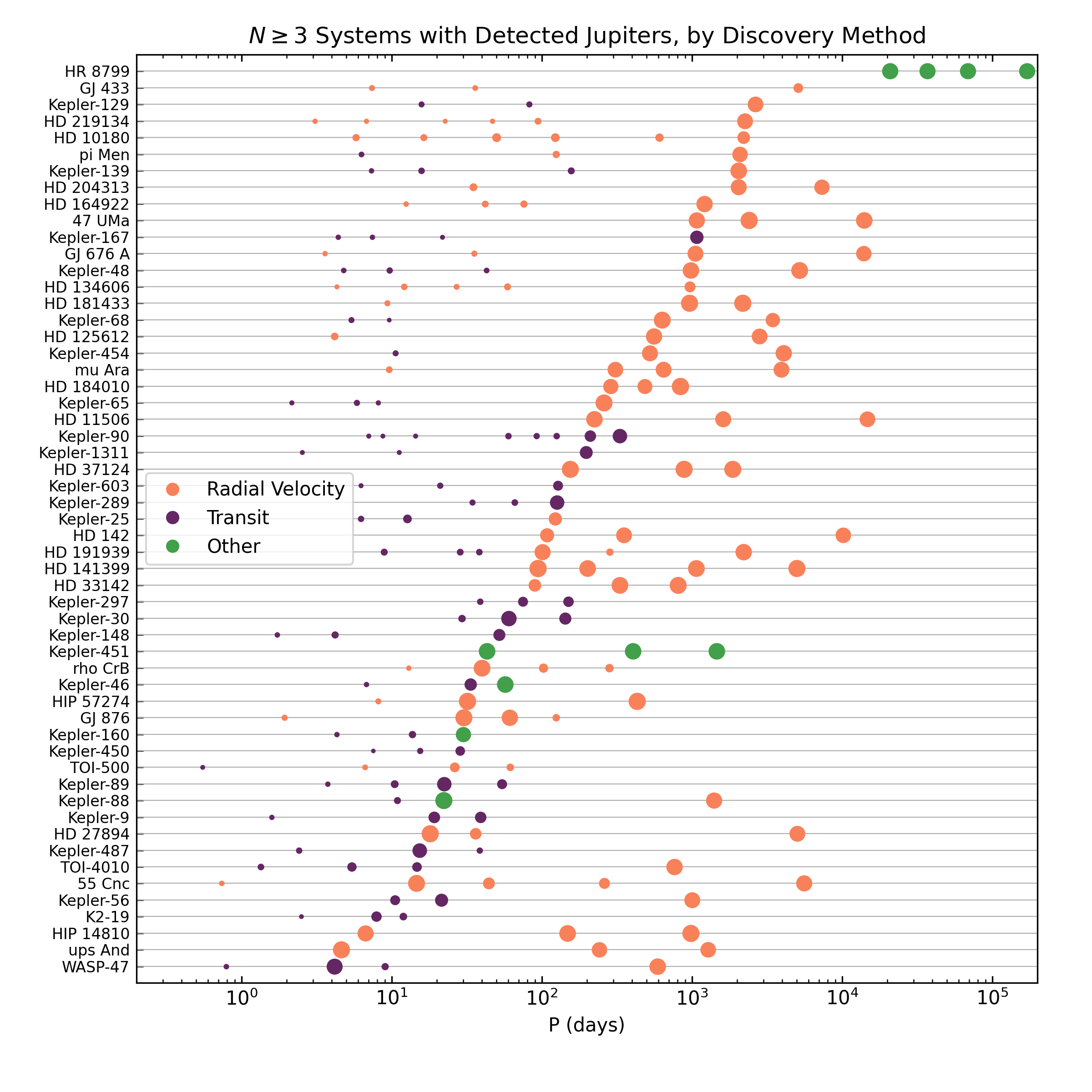}
    \caption{Architectures of \three systems with one or more detected Jupiters (of any period) ordered by period of the first Jupiter in the system and colored by discovery method. Radial velocity discoveries (orange) and transit discoveries (purple) are distinguished from all other detection methods (green). Planets in the same system may or may not have the same discovery method.}
    \label{fig:method}
\end{figure}

The problem of observational biases is of some concern and raises the question of how much of what we see merely derives from artifacts of a biased sample. As an example, we illustrate this issue in Figure \ref{fig:method}, which plots the architectures of a subsample of systems by discovery method. In particular, we plot all \three systems with any number of Jupiters at any periods, distinguishing radial velocities (orange), transits (purple), and other detection methods (green). Notably, transiting planets are much more represented in warm Jupiter systems, while radial velocity detections are biased toward systems with distinct outer planets under our definition. However, this could simply be because transit surveys are more strongly-biased toward short-period planets, and we also note that neither detection method is entirely unrepresented in any part of the parameter space, except the very longest periods, which should bound the possible errors to some degree. Therefore, while observational biases are large confounders on statistics, we believe they are not large enough to negate the \textit{existence} of our defined categories.

Our classification is likewise limited by the size of the available dataset. This paper is based on the notion that the population of observed exoplanets is large enough in aggregate to make such a classification. However, a deeper analysis of e.g. trends with host star spectral type, age, or differences of binary star systems would require an order of magnitude more data, as there would simply not be enough systems to make meaningful categories outside of the main closely-spaced peas-in-a-pod cluster. Thus, the current version of our scheme does not consider the properties of the host star or stars, except to exclude pre-Main Sequence stars and planets that are likely to be ``second generation,'' as discussed in Section \ref{sec:filter}. (In particular, for planets in S-type orbits around the primary of a widely-separated binary system, our classification scheme still applies, although more closely-spaced binaries can truncate the outer edge of an S-type planetary system. As for P-type (circumbinary) planets, too few of them have been detected, especially in multiplanet systems, to draw meaningful conclusions.)

One of the largest limitations of our classification is the difficulty in constructing a rigorous definition of inner and outer planetary systems. As discussed in Section \ref{sec:3plusclass}, our definition is largely empirical, based on the currently known population, and it could change significantly with future discoveries. We see hints of a population or populations of systems that appear transitional with e.g. three Jupiters apparently inside the ice line (HD 33142) or small ``inner'' planets extending out past the ice line (HD 10180). With a larger and more uniform sample of planetary systems, these unusual cases could potentially expand into new categories of their own.

%Especially Jovian-rich inner systems, with three or more Jupiters at short periods, are not detected. This is expected give the improbability of three giant planets either forming or migrating in that close to the star. (The nearest-miss potential example of this class is 55 Cnc.) Further, the few cases that might come close are those that are ambiguous in terms of drawing the line between inner and outer planets, as we discuss in Section \ref{sec:outer}. A larger and more uniform sample of outer plantary systems, such as could better illuminate the inner-outer divide, is likely the largest single enhancement that could be made to our framework.

% FRED: when nature is assembling planetary systems it doesn't just take random planets; there are trends in what we see. Systems are less diverse than is usually suggested in the "everything can happen" sense.

Despite these ambiguities, there is one class of planet and planetary system that is conspicuously absent from our dataset: a system with non-Jovian outer planets -- that is, a non-Jovian planet exterior to a Jupiter that is already classified as an outer planet, or, with a slight relaxation of the definition, a non-Jovian planet separated by a large period gap from an interior Jupiter. (The third possible case, where the \textit{innermost} planet has a period $>$130 days and is a non-Jupiter, is not technically contradicted by any system in our dataset, or even our own Solar System, but it would lead to some very strange result that lead us to reject it in this context. For example, it would classify Venus as an ``outer planet'' if Mercury were not detected from a distance.) 
%Observational biases will be a large part of why we do not see any planets of this type, as smaller planets are much less detectable at periods approaching the ice line. 
%Still, it appears significant that we do not detect any such planets, even in configurations where we potentially could. 
If these omissions are due to the observational biases folded into the exoplanet census, which tend to disfavor the discovery of small, long-period planets, then they may be remedied to some extent by upcoming missions such as the \textit{Roman Space Telescope} \citep{Roman}.

There are potential candidates for non-Jovian outer planets in systems such as Kepler-150 and GJ 163, if the definition of an outer planetary system is relaxed further to include gaps between two non-Jupiters. These systems each have two small planets separated by a large gap, with the exterior planet having a long enough period that it is suggestive of an outer planet. These do not qualify as true outer planets under our definition, and their status could also be revised by the discovery of additional planets in the gaps. However, a larger sample could also reveal a sharper distinction to differentiate them from typical peas-in-a-pod systems.

Uranus and Neptune would also qualify as non-Jupiters in an outer planetary system. However, as discussed in Section \ref{sec:3plusclass}, this is an expected consequence of planet formation at very long periods and they would not impact the delineation of the inner-outer divide at such wide separations.

For inner planetary systems, the parameter space is more fully covered, but there are still categories in our framework that remain empty or nearly empty. Peas-in-a-pod systems are well-represented. Even the rare middle-gap systems are represented by the otherwise normal-looking Kepler-62 (and the less normal-looking HIP 41378). However, warm Jupiter systems, in addition to being fewer in number, have gap structures that are both more ambiguous and harder to fit into our definitions (since a large gap between a non-Jupiter and a Jupiter in most cases will designate outer planets). Warm-Jupiter systems also do not show any middle-gap systems.

Multi-gapped inner systems (that is, those with multiple large period gaps among their inner planets) are also not detected, although there is a near miss in GJ 3138, with period ratios of 4.9 and 43. Explanations for this phenomenon could include the relative unlikelihood of a system being so dynamically sparse, or the larger amount of dynamical space needed to fit multiple large gaps. %(For comparison, for the similarly rare strongly-inverted mass ratios, we would expect to see on the order of 2$-$3 systems with multiple such ``inverted'' pairs of planets. In fact, we see only one, Kepler-20, but this deficit is not statistically significant.)

Finally, there is the as-yet theoretical category of ``anti-ordered'' planetary systems, where planet sizes decrease with distance from the star. This goes against the standard planet formation paradigm \citep{Dai2020,Weiss23}, but such systems are predicted, albeit at low abundance, by some population synthesis models \citep{Mishra22}. No such systems appear in our dataset among \three systems. Although the 2-planet systems Kepler-87 and KOI-1783 would technically qualify, we consider two planets to be insufficient to confirm their anti-ordered character. We have already established that there are no known non-Jovian outer planets under our definition. Meanwhile, for systems with \three inner planets, there are none where the innermost planet is a Jupiter, and only 12 where the innermost planet is even a Neptune, of which four (AU Mic, Kepler-254, Kepler-279, and Kepler-549) could be reasonably considered anti-ordered, with sub-Neptunes strictly exterior to Neptunes. These four do not look dramatically different from other peas-in-a-pod systems, so they do not appear to form a distinct category. This suggests that any anti-ordered class of systems, if it exists, should be required to have inner Jupiters.

These considerations lead us to a shortlist of classes of planetary systems that could apply to future discoveries, but which are not represented in the current dataset, or are too under-represented to clearly delineate:
\begin{itemize}
    \setlength\itemsep{0em}
    \item One or more categories of ``transitional'' systems that do not fit the inner/outer paradigm.
    \item A non-Jupiter that falls in the outer system under our definition.
    \item A coherent category of outer planets in systems without any Jupiters.
    \item Middle-gap warm-Jupiter systems.
    \item Multiply-gapped inner systems.
    \item Anti-ordered systems with all non-Jupiters exterior to all Jupiters.
\end{itemize}

% what does incompleteness mean for us? (see below - notes from last time)
%% Fred: our characterization applies to inner solar systems
% microlensing will fill in gaps (in outer systems) 
% branching ratios will be updated in future with more data  - those are in question, NOT the categories defined above
% these ratios may depend on stellar mass, metallicity, etc as well. 
% could be correction to categories in inner solar system due to ex: Marses, things we haven't been able to detect yet 
% how do categories change with additional planets? dynamically mixed systems will not change category. peas in the pod could have Jupiter/ Saturn location planets that we don't know about and that might 

\section{Discussion}
\label{sec:conclusion}

\subsection{Summary}

This paper presents a classification framework for exoplanet systems that can classify the vast majority of the currently observed population into clearly-delineated categories. Although extrasolar planetary systems are often considered to be highly diverse, this classification scheme indicates that they display a surprising degree of uniformity in their architectures.

This categorization is based on two primary components: (1) the physical locations where planets reside within their planetary systems and (2) the degree of coherency of their orbital and physical parameters. Planetary systems are divided into inner and outer systems based on their orbital periods. In the presence of a wide gap between planets, when the period ratio between two adjacent planets exceeds 5, and the outer planet has a period of $>$130 days, the gap serves as the boundary between the inner and outer system. In cases where there is not such a gap, a system is defined as an outer system if the innermost planet in the system is a Jupiter-sized planet with a orbital period of $>$130 days.

Using our classification scheme, we can summarize the current population of multiple planet systems as follows: After applying our filters for data quality, the database contains 948 multiple systems, with 634 systems containing $N=2$ planets and 314 containing \three planets. The $N=2$ systems show a clear peak of mass ratios $M_2/M_1\approx2$ and period ratios $P_2/P_1\approx3-10$, although the distributions have significant tails (Figures \ref{fig:2planet_p} and \ref{fig:2planet_m}). Within the class of \three systems, a sizable majority (226 out of 314) contain no Jupiters and are closely spaced, and thus display what is usually known as peas-in-a-pod configurations. This number rises to 232 out of 286 when \three \textit{inner} planets are required, and Jupiters that are distant companions (outer planets) are excluded, as in our final classification system. The complete census is listed in Table \ref{tab:multis}.

Despite the diversity in individual planet types, most known inner systems are well-described as one of two main categories: peas-in-a-pod systems or warm Jupiter systems. Peas-in-a-pod systems consist of small, mostly regularly-spaced planets which are typically sub-Neptunes or super-Earths. In contrast, warm Jupiter systems feature Jupiter-mass planets with smaller nearby planets and have more variability in their architectures (e.g., their mass ratios and period ratios). Only a tiny minority of \three systems (9 out of 314) prove difficult to classify into one of these two categories, or as outer planets only. (Systems with hot Jupiters can also be considered as a distinct and mostly-non-overlapping class of inner systems, but the vast majority of them have lower multiplicity.)

In addition, inner systems are classified based on the regularity of their planet spacing. Inner systems can exhibit either (1) closely-spaced orbits (those with approximately even logarithmic spacing in their orbital periods) or (2) irregularly-spaced orbits (those where at least one pair of adjacent planets has a period ratio $>$5). For irregularly-spaced or ``gapped'' systems, our framework provides an additional layer of classification based on the location of the largest gap between inner planets. These gaps typically occur either between the innermost pair of planets or the outermost pair within the inner system.

It is informative to consider the relative abundance of features in systems with different planet types. Out of the 286 systems with \three inner planets, 23 of them contain warm Jupiters. And when looking at irregular spacings among those inner planets (with period ratios $>$5), they occur in 9 out of 23 warm Jupiter systems (39\%), but only 34 out of 266 peas-in-a-pod systems (13\%). This pattern continues for other dynamical properties, with variations in warm Jupiter systems occurring at, very roughly, the 20\%-40\% level, while for peas-in-a-pod systems, they occur at the 5\%-10\% level, highlighting the greater regularity of these systems.

%It is worth noting that no known planetary system exhibits a configuration where an interior Jupiter is followed by an exterior non-Jupiter planet across a large period gap. Similarly, there are no known cases in which a non-Jupiter is the innermost planet in a system of three or more planets with an orbital period exceeding 100 days (except for Mercury in our own solar system). 

\subsection{Implications and Future Work}

This classification framework may be revised or extended in the future in a number of ways. For example, the precise definition of the divide between inner and outer planets can be modified. This definition can evolve as planetary searches are extended to include planetary orbits with larger periods (and hence more outer planets). In addition, specific parts of the parameter space could, in the future, provide evidence of new categories of planetary systems. Future microlensing surveys, such as that planned for the \textit{Roman Space Telescope}, should prove especially useful for filling in these blank corners of the map \citep{Roman}. On another front, the present data do not address binary star systems separately, nor do they include Trojan planets or exomoons; the future discovery of such bodies could necessitate additional categories in the classification scheme.

Our own Solar System, viewed through the lens of the four planets that would be likely to be detected with current methods (Venus, Earth, Jupiter, and Saturn), shows clearly distinguished inner and outer planets,\footnote{\raggedright About 1 out of 6 systems contain Jupiters, as does our Solar System, the vast majority of which have identifiable inner and (where detected) outer planets.} but it lacks the three (visible) inner planets that would be needed to definitively classify it. When all eight planets are included, our Solar System has a closely-spaced peas-in-a-pod inner system, typical of the large majority (80\%) of \three inner systems, albeit with somewhat less uniformity of planet size than average.\footnote{Our inner solar system is less ordered than 4 out 5 peas-in-a-pod systems, but more ordered than 1 out 5 \citep{Weiss23}}. The ``unusually long'' periods of the inner planets would plausibly only appear so because of observational selection effects. Uranus and Neptune do represent non-Jovian outer planets, which do not appear in the current exoplanet catalog, but these may also prove to be common with improved detection techniques. Taken together, results from this classification scheme indicate that while our solar system is somewhat less well-ordered than the mean, it is much less exotic than is often claimed.

One important related issue is that sufficiently small planets, comparable to the mass and size of Mars, can easily escape detection with our current observational capabilities. As a result, relatively small, Mars-sized planets could be hiding within observed multi-planet systems, for example in the inner-outer divide. Although statistical arguments suggest that relatively few {\it detectable} planets are likely to populate the gaps, and many of these gaps are too wide to be filled by a single additional planet, we cannot rule out the presence of sufficiently small bodies. In the future, the presence of such small entities, which act like test particles, can be constrained through dynamical considerations. 

This classification scheme provides a largely qualitative description of the architectures of currently observed multi-planet systems. The next step is to understand how such systems are formed, and, perhaps equally important, why other dynamically plausible systems are not present in the database (see Section \ref{sec:discuss}). Given that peas-in-a-pod systems are the most prevalent, a theoretical understanding of their formation is a high priority. As one part of the explanation, systems with equal spacing can be shown to minimize their energy if they have nearly equal masses, circular orbits, and low mutual inclinations (\citealt{Adams2020,Adams2019}; see also the discussion in \citealt{Weiss23}). These arguments provide only a partial explanation, however, so that the channels by which forming planetary systems reach their minimum energy states remain to be determined. In addition, the most common type of planets in these uniform system have mass of order $10M_\oplus$. In other words, the most common planets in the most common planetary systems are super-Earths, which thus represent a preferred outcome of the planet formation process.

Another open theoretical issue is the rate of formation for Jovian planets, including both their overall occurrence rate and their final locations. This classification scheme suggests that Hot Jupiters are rare ($<$1\% of \three systems) and have few close companions (Figure \ref{fig:HJmultis}); warm Jupiters are relatively uncommon ($\sim$10\%) and often have smaller companions (Figure \ref{fig:warm_jup}); and systems with detected outer planets tend to have Jupiters with periods 300 -- 3000 days (Figure \ref{fig:outer}). These trends are all consistent with previous expectations, with the latter result indicating that the ice-line might be a preferred formation location. The formation rate of Jovian planets is part of the larger question of determining the distribution of planetary masses, i.e., the planetary mass function (PMF). Finding a predictive theory for the PMF is the Holy Grail of planet formation theory. 

Note that the observed architectures of planetary systems have important implications for the prospects of finding habitable planets. On one hand, since most systems have well-ordered peas-in-a-pod configurations, with regularly spaced orbits, most systems will have planets at (very roughly) predictable periods on the order of 10-100 days. For sun-like stars, these distances would be too close for any of the planets to fall in the habitable zone (the right distance from its host star to support liquid water). Around M-dwarfs, however, peas-in-a-pod systems may subtend the habitable zone, allowing for one or more planets to be at appropriate surface temperatures for liquid water. This may suggest that the majority of habitable planet reside around lower-mass stars in peas-in-a-pod systems. 
However, there are unresolved challenges for the habitability of planets around M-dwarfs, due to the stars' XUV irradiation and potential for flaring \citep[see e.g.][]{Tian2018}.
Additionally, as discussed already, most planets in peas-in-a-pod systems are super-Earths, rather than Earth-sized, and may be too large for the canonical definition of a habitable planet. 

From the earliest discoveries of extrasolar planets in the 1990s, the field has faced an increasingly complex zoo of planets and solar systems that were often far different from what was expected. Impressive work has been done to understand this population, yet most empirical analyses have pointed to a broad spectrum with long tails and many exceptions. But just like the puzzle of the particle zoo in the 1960s was solved by the discovery of quarks, the exoplanet zoo also seems to have clear organizing principles that let us classify it into distinct types of solar systems. Though far from complete, we believe this classification provides a better understanding of the population as a whole, and it should be fertile ground for future studies of exoplanet demographics and formation.

\vspace{12pt}

\section*{Acknowledgements}

ARH acknowledges support by NASA under award number 80GSFC24M0006 through the CRESST II cooperative agreement.

ARH and CCS acknowledge support from the GSFC Exoplanet Spectroscopy Technologies (ExoSpec), which is part of the NASA Astrophysics Science Division’s Internal Scientist Funding Model.

ARH thanks Edgar Grunewald of the Artifexian YouTube channel for the inspiration to pursue this line of research.

FCA acknowledges support from the Leinweber Center for Theoretical Physics (LCTP) at the University of Michigan.

This work was performed in part by members of the Virtual Planetary Laboratory Team, a member of the NASA Nexus for Exoplanet System Science, funded via NASA Astrobiology Program Grant No. 80NSSC18K0829.

We thank the anonymous referee whose review of this paper greatly improved the completeness and rigor of our scientific analysis.

This research has made use of the NASA Exoplanet Archive, which is operated by the California Institute of Technology, under contract with the National Aeronautics and Space Administration under the Exoplanet Exploration Program.

This paper makes use of data from the first public release of the WASP data \citep{Butters10} as provided by the WASP consortium and services at the NASA Exoplanet Archive, which is operated by the California Institute of Technology, under contract with the National Aeronautics and Space Administration under the Exoplanet Exploration Program.

This paper makes use of data from the UKIRT microlensing surveys \citep{Shvartzvald17} provided by the UKIRT Microlensing Team and services at the NASA Exoplanet Archive, which is operated by the California Institute of Technology, under contract with the National Aeronautics and Space Administration under the Exoplanet Exploration Program.

This paper makes use of data from the KELT survey, which are made available to the community through the Exoplanet Archive on behalf of the KELT project team.

This paper makes use of data obtained by the MOA collaboration with the 1.8-metre MOA-II telescope at the University of Canterbury Mount John Observatory, Lake Tekapo, New Zealand. The MOA collaboration is supported by JSPS KAKENHI grant and the Royal Society of New Zealand Marsden Fund. These data are made available using services at the NASA Exoplanet Archive, which is operated by the California Institute of Technology, under contract with the National Aeronautics and Space Administration under the Exoplanet Exploration Program.

\bibliography{refs}
\bibliographystyle{aasjournal}   % makes bibtex use spiejour.bst

\end{document}